\documentclass[trackchanges]{aastex631}
\usepackage[T1]{fontenc}
\begin{document}

\title{A Multi-station Meteor Monitoring (M$^3$) System. II. system upgrade and a pathfinder network}
\author[0000-0002-3336-9491]{Zhenye Li}
\author[0000-0002-6684-3997]{Hu Zou}

\affiliation{CAS Key Laboratory of Optical Astronomy, National Astronomical Observatories, Chinese Academy of Sciences, Beijing 100101, P.R. China}
\affiliation{School of Astronomy and Space Science, University of Chinese Academy of Sciences, Beijing 100049, P.R. China}

\author{Jifeng Liu}
\affiliation{New Cornerstone Science Laboratory, National Astronomical Observatories, Chinese Academy of Sciences, Beijing 100101, P.R. China}
\affiliation{School of Astronomy and Space Science, University of Chinese Academy of Sciences, Beijing 100049, P.R. China}
\affiliation{Institute for Frontiers in Astronomy and Astrophysics, Beijing Normal University, Beijing, 102206, P.R. China}

\author{Jun Ma}
\affiliation{CAS Key Laboratory of Optical Astronomy, National Astronomical Observatories, Chinese Academy of Sciences, Beijing 100101, P.R. China}
\affiliation{School of Astronomy and Space Science, University of Chinese Academy of Sciences, Beijing 100049, P.R. China}

\author{Qingyu Meng}
\noaffiliation
\author{Yingjie Cai}
\author{Xinlin Zhao}
\author{Xue Li}
\author{Zhijun Tu}
\author{Bowen Zhang}

\affiliation{CAS Key Laboratory of Optical Astronomy, National Astronomical Observatories, Chinese Academy of Sciences, Beijing 100101, P.R. China}
\affiliation{School of Astronomy and Space Science, University of Chinese Academy of Sciences, Beijing 100049, P.R. China}
\author[0009-0000-9362-3642]{Rui Wang}
\affiliation{CAS Key Laboratory of Optical Astronomy, National Astronomical Observatories, Chinese Academy of Sciences, Beijing 100101, P.R. China}

\author[0009-0007-8133-5249]{Shaohan Wang}
\affiliation{Department of Astronomy, University of Science and Technology of China, Hefei 230026, P.R. China}

\author{Lu Feng}
\affiliation{CAS Key Laboratory of Optical Astronomy, National Astronomical Observatories, Chinese Academy of Sciences, Beijing 100101, P.R. China}

\email{zouhu@nao.cas.cn,jfliu@nao.cas.cn}
\begin{abstract}

Meteors are important phenomenon reflecting many properties of interplanetary dust particles. The study of their origin, mass distribution, and orbit evolution all require large data volume, which can only be obtained using large meteor networks. After meteor networks in Europe and America, we present our designs and upgrades of a proposing network in China. The new designs are mainly aimed for facilitating data gathering process. Each of the newly designed meteor stations now can support up to 4 cameras to cover the full sky. Newer version of meteor station software now works as an integral system, which can streamline the process of detecting, measuring and uploading meteors. We have built a meteor data platform to store, process and display the meteor data automatically. The software and data platform are designed to be easy to learn and use, so it can attract more people to join and operate meteor stations. Four stations are installed as the first phase of the network, and during the operation in 10 months, the network detected 8,683 orbits, and we find that half of the orbits can be related to established meteoroid streams. The statistical analysis of sporadic meteoroids shows a bimodal distribution of the velocities, which coincides with previous studies. The distribution of Tisserand parameters, $T_j$, shows the two peaks at $T_j=0$ and 3, indicating the different orbits of parent bodies (isotropic and ecliptic), which are divided by $T_j=2$. The falling trajectory of a meteorite was also predicted using observational data of the network. We are currently expanding the network, and in the future we will carry out detailed analysis of the key parameters of the distribution of the meteoroids.
\end{abstract}

\section{Introduction} \label{sec:intro}

Meteors are the phenomenon caused by the collision between space particles and Earth's atmosphere \citep{1998Meteor}. The precursors of meteors, i.e. meteoroids, are deeply related with activities and interactions of minor planets of the solar system. Meteor showers, which are caused by comets and active asteroids, can be used to find and study the parent bodies \citep{2011Icar..216...40J}. Furthermore, meteor phenomenon carry unique information about interplanetary environment. 

A continuous monitoring of meteor activities is important for finding new meteoroid streams and possible long-period comet impactors to the Earth \citep{2018pimo.conf...65D,2003Icar..162..443L}. Meteor monitoring is also important for human space activities. Micrometeorites can damage and even disable spacecrafts, e.g. Olympus 1 \citep{DOUGLASCASWELL1995139}. Assessing the threat posed by meteoroid impacts is fundamental in spacecraft design, particularly for those intended for high Earth orbits or embarking on interplanetary missions. Several MEMs (Meteoroid engineering models) have been developed and maintained by the National Aeronautics and Space Administration (NASA) and the European Space Agency (ESA) \citep{2019LPICo2109.6054M,2019A&A...628A.109S}, which provide estimations of meteoroid impact frequencies, speeds, and directions. These models are developed using observational data from radars and optical cameras.

Meteors are detectable through diverse observational means, encompassing optical, radar, and infrasonic wavelengths, each providing unique insights into these celestial phenomena \citep{1998Meteor}. During and after their atmospheric entry, the glowing trails are observable from the ground. In the optical wavelength, the observation of a meteor from multiple distinct locations permits the determination of its 3D trajectory. This, in turn, facilitates the derivation of the original orbit in the solar system and, potentially, the identification of the impact location of any surviving meteoritic fragments \citep{1998Meteor,2020MNRAS.491.2688V}. Consequently, meteor phenomena are systematically monitored by meteor observation networks, comprised of a multitude of geographically dispersed stations. At each station, video cameras monitor the sky and record videos of meteors, and then extract information of the meteoroid movement in order to calculate the 3-D trajectories.

To accurately determine the 3D positions of meteoroids, a baseline spanning tens of kilometers between observing stations is crucial. Moreover, given the sporadic nature of meteor occurrences, broad ground coverage is vital for effective meteor monitoring. The European Fireball Network is the first to adopt this method dating back to 1950s \citep{1987BAICz..38..222C, 2022A&A...667A.157B}. Its pioneering efforts inspired the development of hardware and data analysis methods. Nowadays, networks with numerous stations and extensive geographic reach are thereby advantageous, as they enable the collection of a rich dataset of meteor events. Notably, there are several extensive meteor monitoring networks globally, such as the Global Meteor Network \citep[GMN;][]{2021GMN}, the Fireball Recovery and Interplanetary Observation Network \citep[FRIPON;][]{2020A&A...644A..53C}, and the Australian Desert Fireball Network \citep[DFN;][]{2017How}. All these networks use CMOS cameras for monitoring meteors. 

In China, there is no meteor network that is as extensive as the ones in Europe and America. However, a large number of meteor cameras are operated mainly by amateurs. In spite of the use of low-cost CMOS cameras similar to those in Europe and America, the software used by the amateurs is not satisfied. Video streams from the cameras are relayed by a streaming software OBS (Open Broadcaster Software) before feeding into a meteor detection software UFOCapture \footnote{\url{http://sonotaco.com/soft/UFO2/help/english/index.html}}. This software is a commercial software developed by Sonotaco \citep{2005pimo.conf..104Y}, and widely used by many network around the world. However, the combination of inappropriate hardware and relaying OBS software resulted in significant degrade of image quality and timing precision.

To determine the orbit of a meteoroid, at least two observations from different locations are required. So naturally the data from different locations need to be collected to perform all the calculations. In other networks around the world, the data are transferred physically (swapping hard drives or mailing DVDs) or through Internet. In China, the operation of stations lacks coordination because of the absence of an automated pipeline. As a result, very few meteors' orbits are obtained.

We try to mitigate all these difficulties by designing new meteor detection and processing system (M$^3$), which is described in previous paper \citep{2024li}. Our system can position the meteors at arc-minute accuracy and calculate trajectories at accuracy of 10 meters. Additionally, our meteor stations have GPS modules to put hardware timestamp on each frames which has precision of 1$\mu$s as tested by \cite{2023PASP..135b5001K}, and uncompressed video data can be saved for further analyses. We also try to get the detailed meteoroid orbits and the photometric data from a bigger sample volume obtained by a larger network. The distributions of speed, orbit elements and bulk densities can be used for the refinement of current meteoroid engineering models \citep{2019LPICo2109.6054M,EHLERT2020249,2011A&A...530A.113K}.

This paper will discuss the upgrade of the system which can be suitable for a bigger network. Section \ref{sec:software} describes the architecture of newer meteor monitoring software. Section \ref{sec:platform} introduces the meteor data platform that processes and stores meteor data automatically. Section \ref{sec:hardware} introduces the modification of the original optical designs. A pathfinder network was established in the north of China, and preliminary results are shown in Section \ref{sec:results}. The summary is in Section \ref{sec:summary}.

\section{Meteor station software}

\label{sec:software}
Meteor station software plays a crucial role in our meteor monitoring network. The software controls the cameras at a station, detects meteors from the video stream, and measures coordinates of detected objects. It is also responsible for communicating with the central server, listening for control commands, and uploading scientific results. Additionally, the software has the capability to run continuously for long periods with minimal maintenance.

We have developed a comprehensive client software that handles all of the above functions for our meteor monitoring stations \citep{2024li}. The client software consists of several components:
\begin{itemize}
\item \textit{meteorThread}. This controls cameras and detects meteors.
\item \textit{meteorExtractor}. This measures coordinates of meteors, as well as generates reports and data files.
\item \textit{meteorWorker}. This schedules operations for the above components, handles their control, and recovers from crashes.
\item \textit{meteorNodeClient}. This maintains WebSocket connections with the central server, and uploads data files.
\item A GUI (Graphical User Interface) component displays basic operational status locally.
\end{itemize}

\begin{figure}[!htb]
    \centering
    \includegraphics[width=1.0\textwidth,trim=1cm 7cm 2cm 5cm,clip]{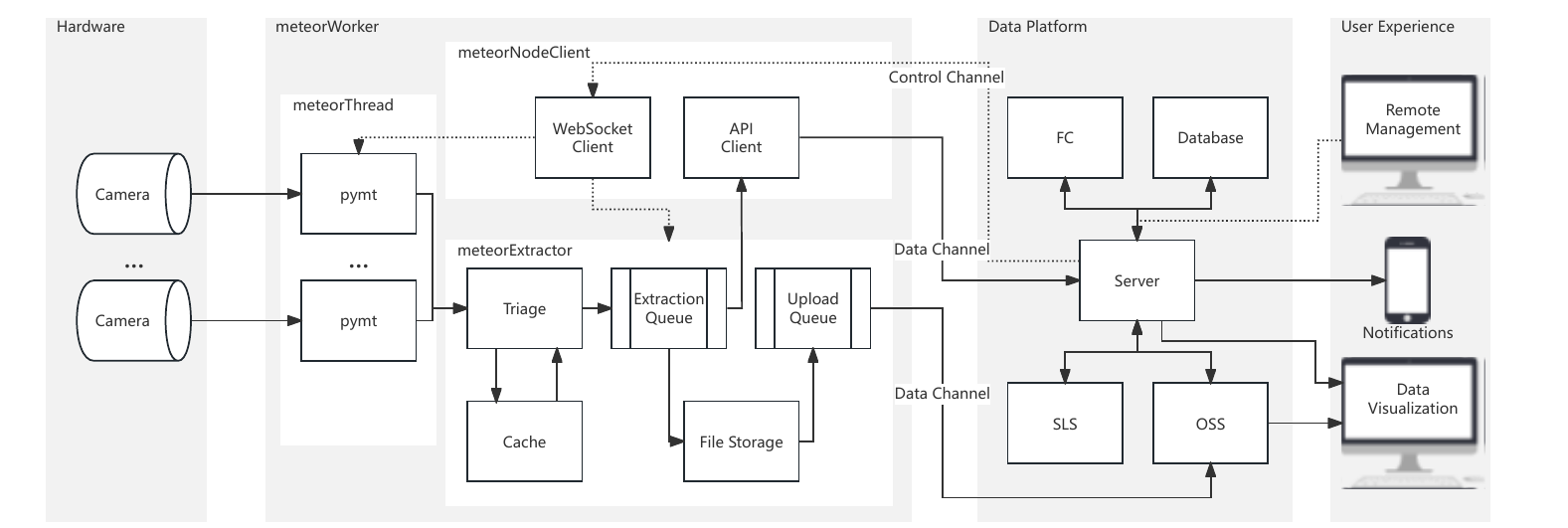}
    \caption{Basic structure of the meteor monitoring system. At each station, \textit{meteorWorker} handles the scientific pipeline in general and communicates with the server using \textit{meteorNodeClient}. Within \textit{meteorWorker}, there is one \textit{meteorThread} process corresponding to each camera. Python objects are passed to \textit{meteorExtractor} for detailed processing, which has only one process. The stations connected to the data platform for controlling the operation and transferring the data. The data platform handles orbit determination and data storage using various cloud services, and provides user interfaces for data visualization and remote control.}
    \label{fig:software}
\end{figure}

The details of the algorithms of \textit{meteorThread} and \textit{meteorExtractor} have been discussed in the previous paper \citep{2024li}. The architecture of our client is shown in Fig. \ref{fig:software}. And we will discuss this further in the following subsections.

\subsection{Scientific pipeline}
Two main components of the scientific pipeline, responsible for detecting and measuring meteors, are \textit{meteorThread} and \textit{meteorExtractor}, which were described in the previous paper \citep{2024li}. The \textit{meteorThread} component detects moving objects within the video feed and generates data which containing video clips and basic object information. These data are then sent to \textit{meteorExtractor} for more detailed processing, during which the coordinates of objects are measured.

The \textit{meteorThread} component, which has stringent requirements for real-time image processing, and it is written in C++. Meanwhile, \textit{meteorExtractor} is written in Python to incorporate several widely used libraries. In the previous version, there are two separate software components, with the data being transferred as FITS files \citep{2010A&A...524A..42P}. The process of the previous version involved writing and reading the file between the two steps, which slowed down the rate of data transfer. In the latest version, \textit{meteorThread} has been compiled as a Python library that can be imported into a Python script. As a result, the data structure of each meteor can be directly retrieved by \textit{meteorExtractor} as Python objects in the computer memory instead of writing and reading files. The Python objects that represent each meteor contain raw video frames, as well as frames depicting temporal maximum, average, and standard deviation (std). Additionally, they record timestamps for each frame, the location of the camera, and other utility information. The details of the data structure will be described in Section \ref{chap:data}.

Each instance of the two software components runs in separate processes that are spawned by \textit{meteorWorker}. At each station, there are multiple \textit{meteorThread} processes, and each process controls a camera. Most of the CPU computation is allocated to \textit{meteorThread}, and in contrast, there is only one process for \textit{meteorExtractor}. Meteor objects are processed as a queue in memory. If the object queue becomes too long, some objects are saved to disk until the queue is empty.  Each process has its own watchdog thread that can reduce and recover from crashes, which has been proven to be vital during long-term operations. Additionally, each \textit{MeteorThread} process has a scheduler to enable automatic starting and stopping of the detection based on the sunrise and sunset times. At the end of daily operation, the schedule is refreshed.

The \textit{meteorWorker} process is responsible for maintaining the complete configuration of the entire meteor monitoring station, which includes the detection schedule, camera settings, detection parameters of \textit{MeteorThread}, measurement parameters of \textit{MeteorExtractor}, camera pointing and distortion models, and file saving and uploading policies.

\subsection{Data Structures}
\label{chap:data}
Data structures have been developed to facilitate the transfer, processing, and storage of the data of meteors and other objects. These structures manifest in three distinct forms: Python objects, files, and database entries. They can be converted into other formats to enhance data transfer and exchange.

Python objects are generated either by \textit{meteorThread} or by reading a FITS file. Each Python object includes an uncompressed video segment, timestamps for each frame, the average frame, std frame, and the binary frame outlining the objects and masked area. Following the processing, the object incorporates the pixel and celestial coordinates of the meteoroid on each frame.

Various formats are employed for saving and displaying meteor data, with the primary form being the FITS file, which can be converted from a Python object. The operations of FITS files are facilitated by \textit{Astropy} \citep{2022ApJ...935..167A}. In this format, image data are stored in the data section, while timing and coordinate data reside in the header, forming header and data units (HDUs). The overall information is stored in the first HDU. Subsequent HDUs represent each frame of the video clip, with timestamps in the header. After processing, the last HDU is appended as a table HDU with pixel and celestial coordinates of the stars in the table, and camera distortion and orientation parameters in the header. Additional image (in PNG format) and video (in MP4 format) formats can be generated for display. In these image formats,  stars and objects are labeled, and additional detailed information is incorporated into the images as text files in JSON (JavaScript Object Notation) format.

The database retains all information except images for efficient querying. Entries can be described as a JSON string, ensuring swift processing, as only coordinate data are essential for orbit determination.

The details of the data structure are listed in Table \ref{tab:data}, in which we label out the included data content of each format with check marks. JSON format covers the hardware information and coordinates without image data, while PNG and MP4 formats cover mostly image data. FITS format includes all of the image, information and coordinate data.

\begin{table*}[htb!]
\centering
\caption{Contents of different data formats}
\label{tab:data}
\begin{tabular}{@{}lllll}
\hline
\hline
Category & Item & FITS & JSON & PNG and MP4\\
\hline
hardware&station code&\checkmark&\checkmark&\checkmark\\
&camera code&\checkmark&\checkmark&\checkmark\\
&station location&\checkmark&\checkmark&\\
&color mode&\checkmark&\checkmark&\\
&frame width&\checkmark&\checkmark&\\
&frame height&\checkmark&\checkmark&\\
&processed time&\checkmark&\checkmark&\\
image data&raw video&\checkmark&&\\
&compressed video&&&\checkmark\\
&average frame&\checkmark&&\\
&std frame&\checkmark&&\\
&maximum frame&\checkmark&&\checkmark\\
&mask frame&\checkmark&&\checkmark\\
&object outline frame&\checkmark&&\\
&frame counts&\checkmark&&\checkmark\\
timing&timestamps&\checkmark&\checkmark&\\
&time data source&\checkmark&\checkmark&\\
stars&star pixel coordinate&\checkmark&\checkmark&\checkmark\\
&star HIP number&\checkmark&\checkmark&\checkmark\\
&star celestial coordinate&\checkmark&&\\
camera model&camera distortion model&\checkmark&\checkmark&\\
&camera rotation matrix&\checkmark&\checkmark&\\
&camera model residual&\checkmark&\checkmark&\checkmark\\
object&object code&\checkmark&\checkmark&\\
&object type&\checkmark&\checkmark&\\
&object pixel coordinates&\checkmark&\checkmark&\checkmark\\
&object celestial coordinates&\checkmark&\checkmark&\\
\hline
\end{tabular}
\end{table*}

\subsection{Communication and Control}
Meteor monitoring stations require continuous communication with the central server. Data flow for these stations can be categorized into two types: engineering data and scientific data. These types of data are transferred via two different channels as follows:

\begin{itemize}
\item Engineering data are transmitted through the WebSocket protocol. This channel is responsible for dispatching configurations and commands to each station, collecting status reports, which includes CPU and memory usage, storage space, process status, and new meteor detections. These reports are concise, but demand a reliable network connection, which is the reason we choose WebSocket.

\item Scientific data are uploaded using representational state transfer (REST) requests. These encompasses files such as JSON, FITS, PNG, and MP4, generated by \textit{meteorExtractor}. To minimize the cost of the server bandwidth, large image files are uploaded to cloud storage, while only text reports are stored in the database of the central server. Upon meteor detection, JSON format reports are initially uploaded to the server, and then pre-signed uploading URLs for file uploads to cloud storage are returned. Subsequently, all remaining files are uploaded in a queued manner.
\end{itemize}

\subsection{Other Functionalities}
In addition to the essential functionalities provided by the software, the client also incorporates additional features to meet specific user needs. These functionalities are as follows:

\begin{itemize}
\item The augmented reality (AR) streaming function of \textit{meteorThread} offers real-time video display for streaming. When provided with camera pointing and distortion parameters, the function layers a star chart on the video. The layer can also highlight the positions of detected meteors one second earlier than they appear in the streaming by caching frames for approximately one second, as depicted in Fig. \ref{fig:ar}. This function is particularly useful for educational streaming during meteor showers, allowing viewers to learn about the sky and receive advance notification of meteor appearances.

\begin{figure}[!htb]
    \centering
    \includegraphics[width=1.0\textwidth]{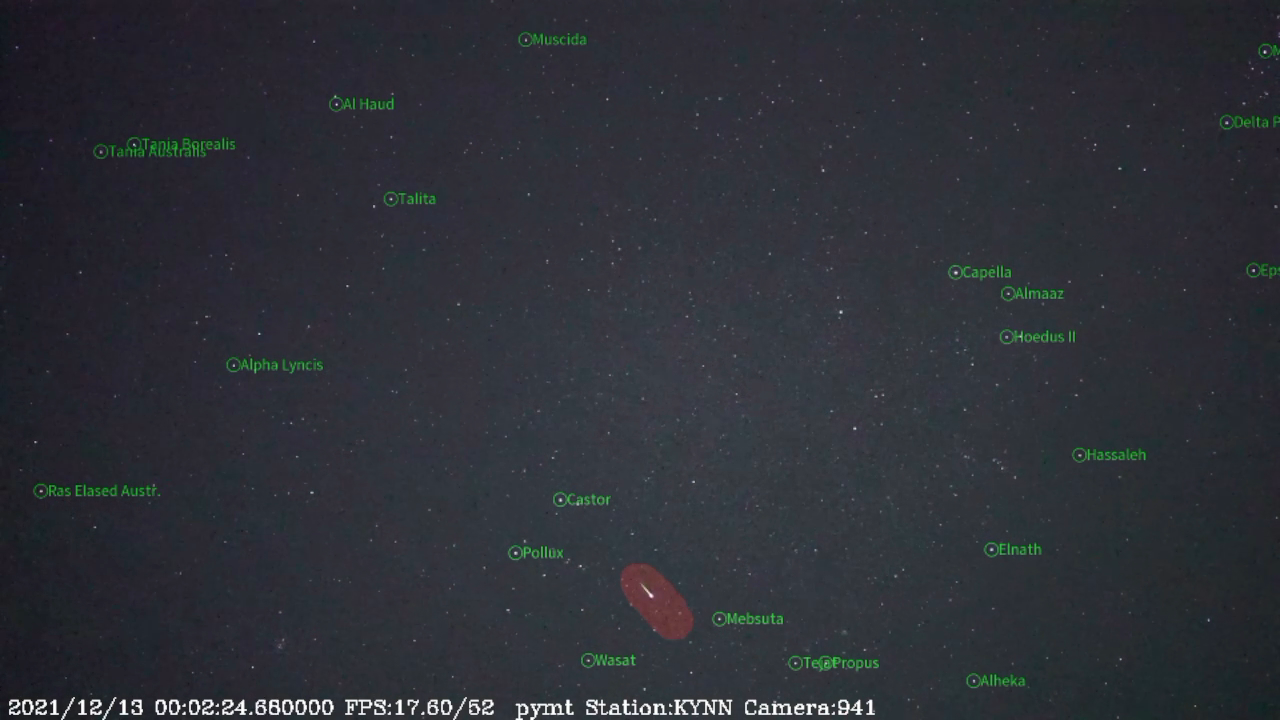}
    \caption{Screenshot of the AR videos as the demonstration of a educational livestream, which is captured by a SONY mirrorless camera at a deeper limiting magnitude. In the video, bright stars are labeled with green circles and their names. The meteor about to appear is highlighted with a red blob. 
    }
    \label{fig:ar}
\end{figure}

\item ADS-B (Automatic Dependent Surveillance–Broadcast) recording function can match detected objects with aircrafts. ADS-B is a technology used by aircrafts to broadcast their positions and other information. Using an additional software defined radio (SDR) module and an antenna, the meteor station can receive and record ADS-B signals from passing aircrafts. This signals are then processed and recorded by the \textit{pyModeS}\footnote{\url{https://github.com/junzis/pyModeS}} which is a Python decoder for Mode S and ADS-B signals. When slow-moving objects are detected using the \textit{meteorThread}, \textit{meteorExtractor} can query records to determine if there are corresponding aircrafts. This function provides the stations with the capability to monitor air traffic.

\end{itemize}

\section{Meteor data platform}
\label{sec:platform}
The meteor data platform is a software system responsible for managing and operating server-side functionalities. Its principal function is to calculate the three-dimensional trajectories of meteoroids. In addition to this, it is also responsible for transmitting user commands to the stations, tracking the network status, managing video data storage, and visualizing data.

We implemented the Software as a Service (SaaS) model for both the station software and data platform. Owners of the stations start the software using a token granted by the data platform. After logging in, users can remotely access the station software through the web-based user interface. The meteor data generated by each station, in accordance with the user preferences, is uploaded to the data platform automatically. This architecture maximizes the efficiency of installing and operating meteor stations for multi-station observation.

The basic structure of the platform and its relationship with station hardware and users are shown in Fig. \ref{fig:software}. Details of the user interface will be described in subsection \ref{sub:UI}, and underlying cloud technologies will be discussed in subsection \ref{sub:back}.

\subsection{User Interface}
\label{sub:UI}

The website user interface is implemented using \textit{React.js}, and it is responsible for all user interactions. The website manages the user authentication, registration, and invitation. During the early stage of the network, only users with valid invitation codes obtained from other users or administrators can access the network. These invitations can be created on the website, and new users can register by accessing the registration page through the invitation link.

Additionally, the website facilitates remote station management by communicating with the station software through a WebSocket connection, providing users with a step-by-step method to configure new cameras. The steps are listed below:

\begin{itemize}
    \item Camera parameters such as the name of the hardware, exposure settings, gain levels, etc, are selected.
    \item The website sends a command to the station to capture and upload an image, allowing the user to draw a mask to exclude non-sky regions in the field of view (FOV).
    \item Upon specifying the basic parameters and the mask, the station is instructed to carry out a comprehensive evaluation of the camera’s performance, including its stability, noise levels, and most notably the full width at half maximum (FWHM) of the star point spread function (PSF). The final configurations are then derived from these test reports. The area threshold for detecting meteors is set to twice the area of the stars, defined as 2$\pi\times$FWHM$^2$.
    \item When the weather conditions permit, the user has the option to manually calibrate the camera. The camera pointing and distortion model values can be estimated by resolving the sample frame via Astrometry.net (\url{https://nova.astrometry.net/}).
\end{itemize}

Following the initial setup, the camera is then able to begin regular operations. Users may fine-tune the camera configurations through the website. In addition, by providing access to another user, they can collaborate and support each other during both initial configuration and daily maintenance.

\subsection{Cloud Based Backend}
\label{sub:back}

The platform is constructed using various cloud services, primarily supplied by \textit{Aliyun}. Rather than running on a single server, the essential functions of the data platform are facilitated by different cloud products. The cost is proportional to the amount of usage, and the cloud services offer exceptional scalability, unchained from the performance limitations of a single machine. The cloud products employed in the platform include:

\begin{itemize}
    \item The Object Storage Service (OSS) is employed for storing a wide variety of files, including meteor files that are organized based on their corresponding station ID. Authorized users can access these files through their pre-signed uploading URLs. Additionally, the static files of the front-end website is also serviced by the OSS.
    \item The Function Compute (FC) is employed for computing meteoroid orbits. It initiates computation instances for each job submitted by the data platform, and subsequently destroys them upon completion. The computation instances calculate the orbits of the meteoroids using \textit{meteorStitch} \citep{2024li}, which is developed based on the Python module \textit{WesternMeteorPyLib} \cite[WMPL;][]{2020MNRAS.491.2688V}. This module employs the model described in \cite{10.1093/mnras/sty1841} to estimate the velocity loss prior to thermal ablation and to derive the meteoroids' velocity at infinity.
    \item The Simple Log Service (SLS) is a cloud-based service utilized for logging all aspects of the data platform. It furnishes widespread monitoring of platform activities and serves to diagnose issues encountered. 
    
\end{itemize}

\section{Meteor station hardware}

\label{sec:hardware}
\subsection{Hardware Upgrades}

\begin{figure}[!htb]
    \centering
    \includegraphics[width=0.5\textwidth]{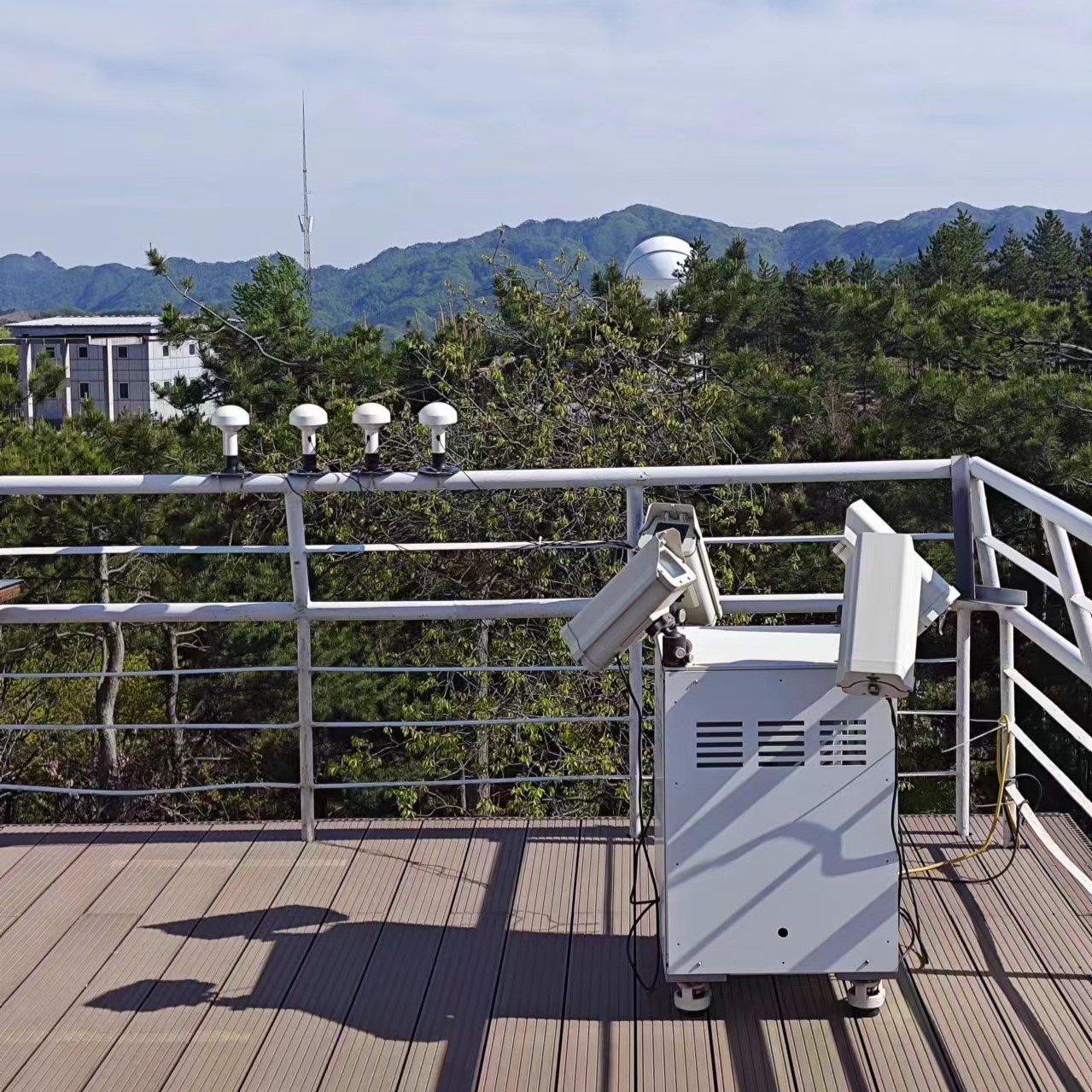}
    \caption{Station hardware installed on the roof of 60/90cm Schmidt telescope at Xinglong station, as an example of exterior designs of the stations.
    }
    \label{fig:station}
\end{figure}

Our station hardware was upgraded to ensure the construction of a large network and better performance. The upgrades are listed in Table \ref{spec_table}. The new design now can support up to 4 camera compartments with different orientations, cameras, and lens combinations. The camera compartments can pivot separately to cover different parts of the sky. To support the simultaneous operation of 4 cameras, a much bigger instrument compartment was designed comparing to the hardware in the previous paper \citep{2024li}, which houses a more powerful computer controlling all 4 cameras. Fig. \ref{fig:station} shows the station hardware installed on the roof of the 60/90cm Schmidt telescope at Xinglong station, as an example of exterior designs of the stations.

\begin{figure}[!htb]
    \centering
    \includegraphics[width=0.8\textwidth]{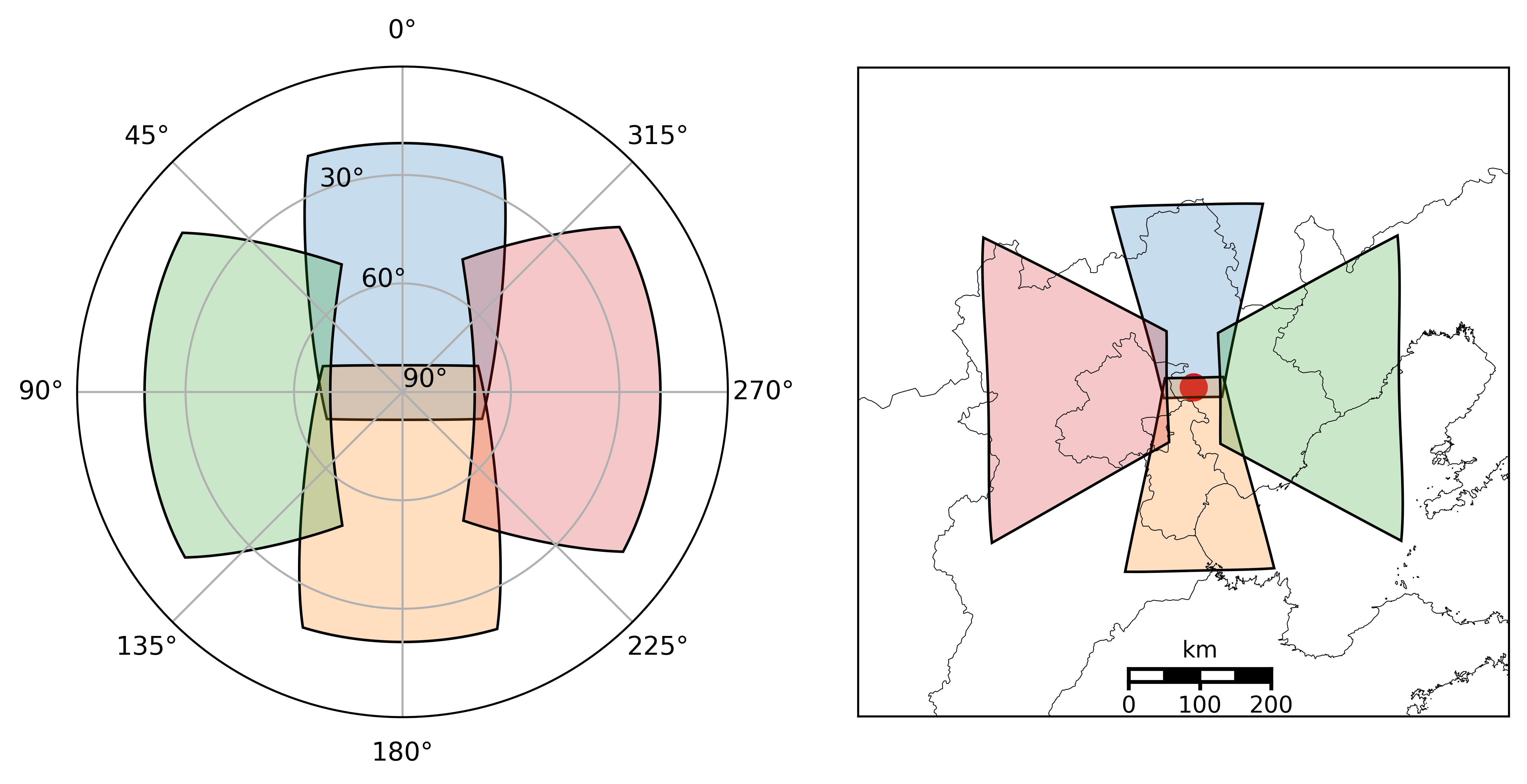}
    \caption{Sky coverage (left) and ground coverage (right) of a meteor station. Each of the 4 cameras is equipped with a 8mm lens.
    }
    \label{fig:cover}
\end{figure}

For the new meteor station, 8mm lenses are mounted on all cameras. This lens has a longer focal length than the 6.5mm lens used in the previous design, thus it has a narrower field of view (FOV) but a larger aperture. With this lens, a tiled arrangement for the camera FOVs can be achieved, covering most of the overhead sky above 20$^\circ$ elevation, and ground area of about 200,000 km$^2$, which is shown in Fig. \ref{fig:cover}.

The brightest meteors have the potential for meteorite recoveries, but their extreme brightness often saturates CMOS detectors, making it difficult to pinpoint their locations. To address this problem, we also tested a much wider 5mm lens at some stations, optimized for accurately detecting bright meteors. This lens can cover a larger area but with a smaller aperture.

The settings of the cameras and software are standardized. All cameras operate in 8-bit mode to reduce the size of image data. The gain of the cameras is changed to 361 from 480 (shown previously as 48 in \cite{2024li} due to different definitions in different software versions, i.e. ten times 48), providing sufficient dynamic range to detect faint objects without being saturated by the brightest stars. A laboratory test conducted with these settings showed that the gain value is 2.2 e$^{-}$/ADU and the readout noise is 3.8 e$^{-}$.

\begin{table*}[htb!]
\centering
\caption{Upgrade of the M$^3$ system.}
\label{spec_table}
\begin{tabular}{@{}lll}
\hline
\hline
Item & Original & Upgraded\\
\hline
Number of cameras&1&4\\
Lens&AZURE-0614MLM &KM0814MP8\\
Fov&$88^\circ\times58^\circ$& $77^\circ\times51^\circ$ \\
Focal length&6.5mm&8.0mm\\
Gain setting&480&361\\
CPU&intel i5-8250U&intel i5-12500H\\
RAM&8GB&16GB\\
Storage&120GB&480GB\\
Power comsumption&30W&50W\\

\hline
\end{tabular}
\end{table*}

\subsection{Station Deployment}

\begin{figure}[!htb]
    \centering
    \includegraphics[width=0.5\textwidth]{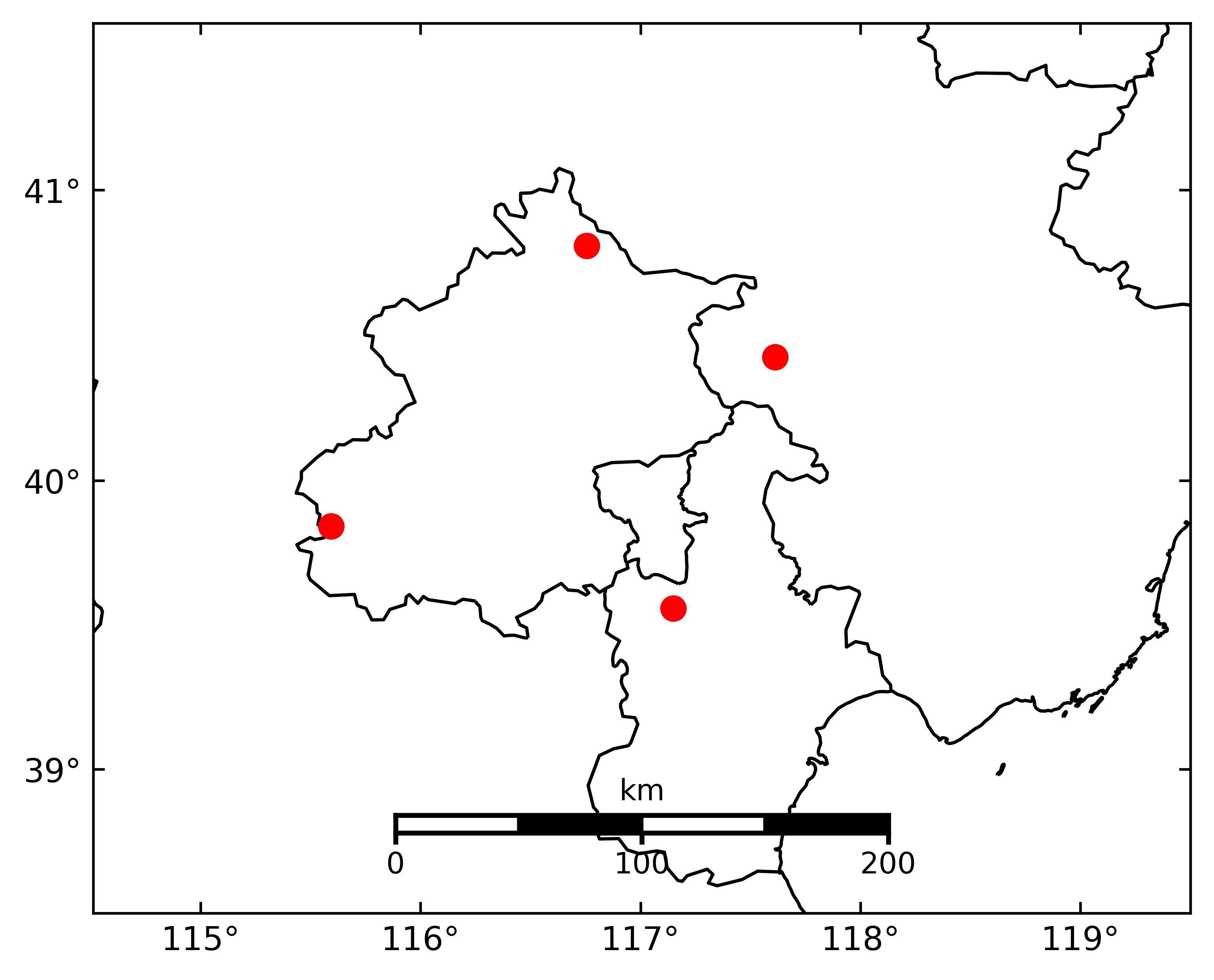}
    \caption{The positions of four stations around Beijing city. All the stations are installed at locations with relatively low light pollution.
    }
    \label{fig:map}
\end{figure}

In the first half of 2023, four stations are deployed around Beijing City. The stations are placed 100-200 km away from each other and roughly form a diamond shape which is shown in Fig. \ref{fig:map}. The locations of the stations are selected based on these factors: power and network support, maintenance, and avoidance of light pollution. Of the stations, two stations (Xinglong and Wuqing) are located at the observational stations of the National Astronomical Observatories, Chinese Academy of Sciences. Baihuashan station is in a natural reserve, and Changshaoying is at an elementary school in the north of Beijing. All the stations are equipped with 4 cameras, pointing in four directions of the sky. Considering ground obstructions, the overall overlap area suitable for triangulating meteors at a 100 km height is 267,800 km$^2$.

\begin{figure}[!htb]
    \centering
    \includegraphics[width=0.5\textwidth]{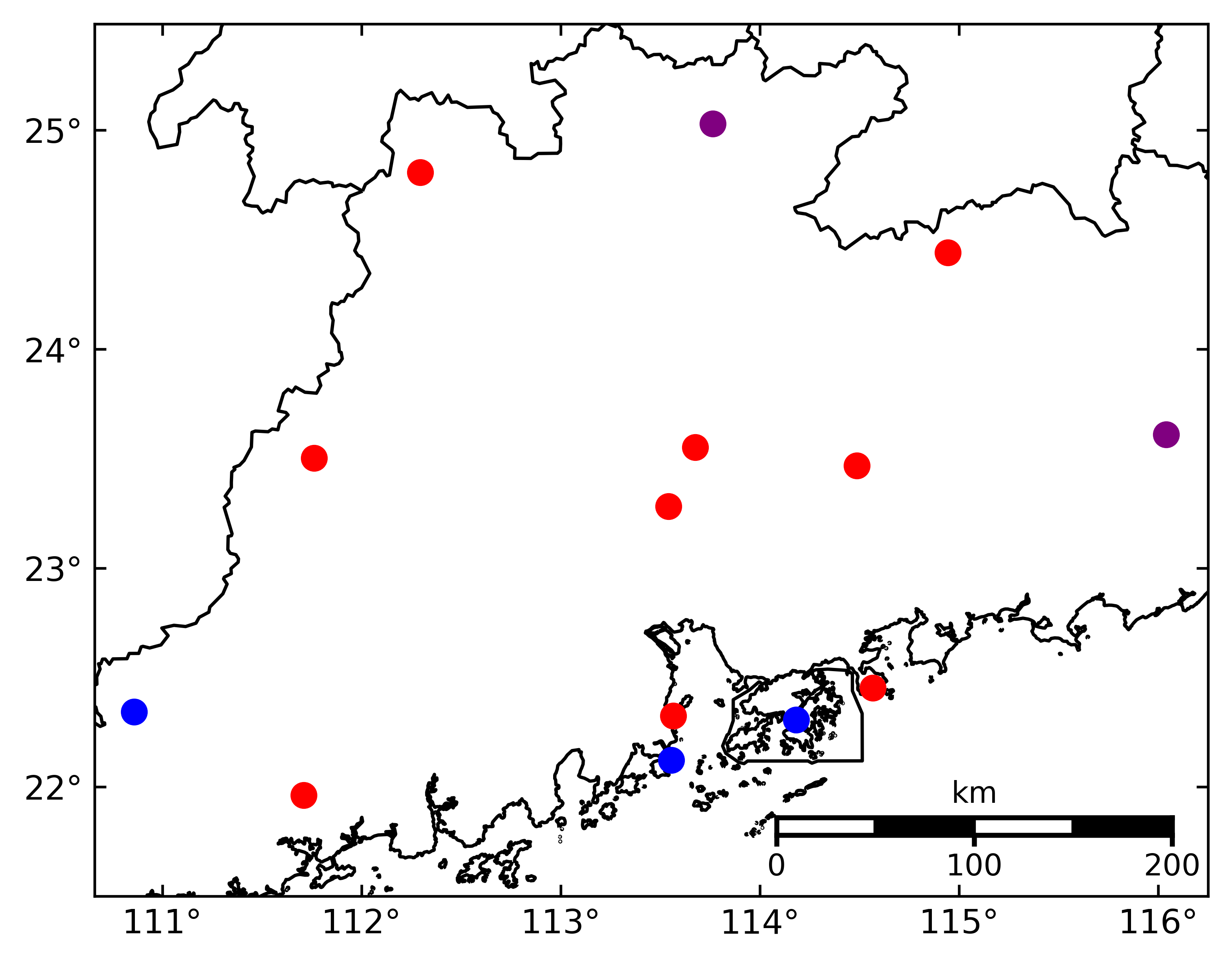}
    \caption{The positions of 14 stations in Guangdong province, Hong Kong and Macau. Red labels stand for standard stations with multiple 8mm lenses, and blue labels stand for stations with a single 5mm lens. Purple labels stand for stations equipped with both types of lenses.
    }
    \label{fig:gdmap}
\end{figure}

In December 2023, another 14 stations are deployed in the Guangdong province and neighboring Hong Kong and Macau, which is shown in Fig. \ref{fig:gdmap}. Most stations are also equipped with similar 8mm lenses, and 5 of the stations have additional 5mm lenses for wider coverage for bright meteor events, as labeled by different colors in the figure. The network has started collecting data from the end of 2023.

\section{Results}
\label{sec:results}

\subsection{Upgrades of Optics}

The lenses with a bigger aperture can result in an improvement in positional precision. Using the same sensors, longer focal length and narrower FOV bring higher angular resolution. On the other hand, a bigger aperture increases the number of photons collected, making it possible to detect fainter meteors.

Before the deployment of the permenant stations of the network, we arranged a side-by-side comparison during the 2021 Geminids meteor shower to compare the performance of the two lenses. Two temporary stations were installed at Yanqihu and Yanqing with a distance of 55 km. At each station, two cameras were installed and pointed in the same direction, one with a 6.5mm lens and the other with an 8mm lens. Meteoroid orbits measured by the two stations are compared, including the detection number and astrometric accuracy. 

During the 3-night observational period, the networks with 6.5mm and 8mm lenses obtained 1,469 and 1,863 single-station meteors, respectively. This shows that despite the narrower FOV, 8mm lenses can detect more meteors due to the deeper limiting magnitudes. However, due to the smaller intersection area of the coverage, the network with 8mm lenses obtained a similar number of meteoroid orbits as the network with 6.5mm lenses, i.e. the latter and the former detected 396 and 416 orbits, respectively.

We also compared the angular precision between the 6.5mm and 8mm lenses using a similar method adopted in \cite{2024li}. \textit{meteorExtractor} performs online calibration for each camera and meteor \citep{2024li}, utilizing a model based on the Brown-Conrady distortion model \citep{1971Close}. This process is described by Equations \ref{equ1} and \ref{equ2}. The parameters of the distortion model include the focal length ($f$), the position of the optical center ($x_0$, $y_0$), and two polynomial coefficients ($k_1$, $k_2$) that characterize the non-linear distortion \citep{2024li}.

When transforming the pixel coordinates ($x_i$, $y_i$) to celestial coordinates ($x_{ci}$, $y_{ci}$, $z_{ci}$), the distances in pixels from the optical center ($x_i - x_0$, $y_i - y_0$) are corrected using a polynomial defined by the two coefficients. Together with the focal length ($f$), the Cartesian coordinates ($x_i^{\prime}$, $y_i^{\prime}$, $f$) relative to the camera can be obtained, which are then transformed into celestial coordinates using a rotation matrix ($\mathbf{R}$). These parameters are initially fitted using known stars in the images, allowing the celestial coordinates of the meteors to be subsequently obtained.

\begin{equation}
\left(
\begin{array}{l}
x^\prime_{i} \\
y^\prime_{i}
\end{array}
\right)
=
\left(
\begin{array}{l}
x_i-x_0 \\
y_i-y_0
\end{array}
\right)
(1+k_1r^2+k_2r^4+\cdots), \label{equ1}
\end{equation}

\begin{equation}
\left(
\begin{array}{l}
x_{ci} \\
y_{ci} \\
z_{ci}
\end{array}
\right)
=\mathbf{R}
\left(
\begin{array}{l}
x_i^{\prime} \\
y_i^{\prime} \\
f
\end{array}
\right), \label{equ2}
\end{equation}

\begin{figure}[!htb]
    \centering
    \includegraphics[width=0.9\textwidth]{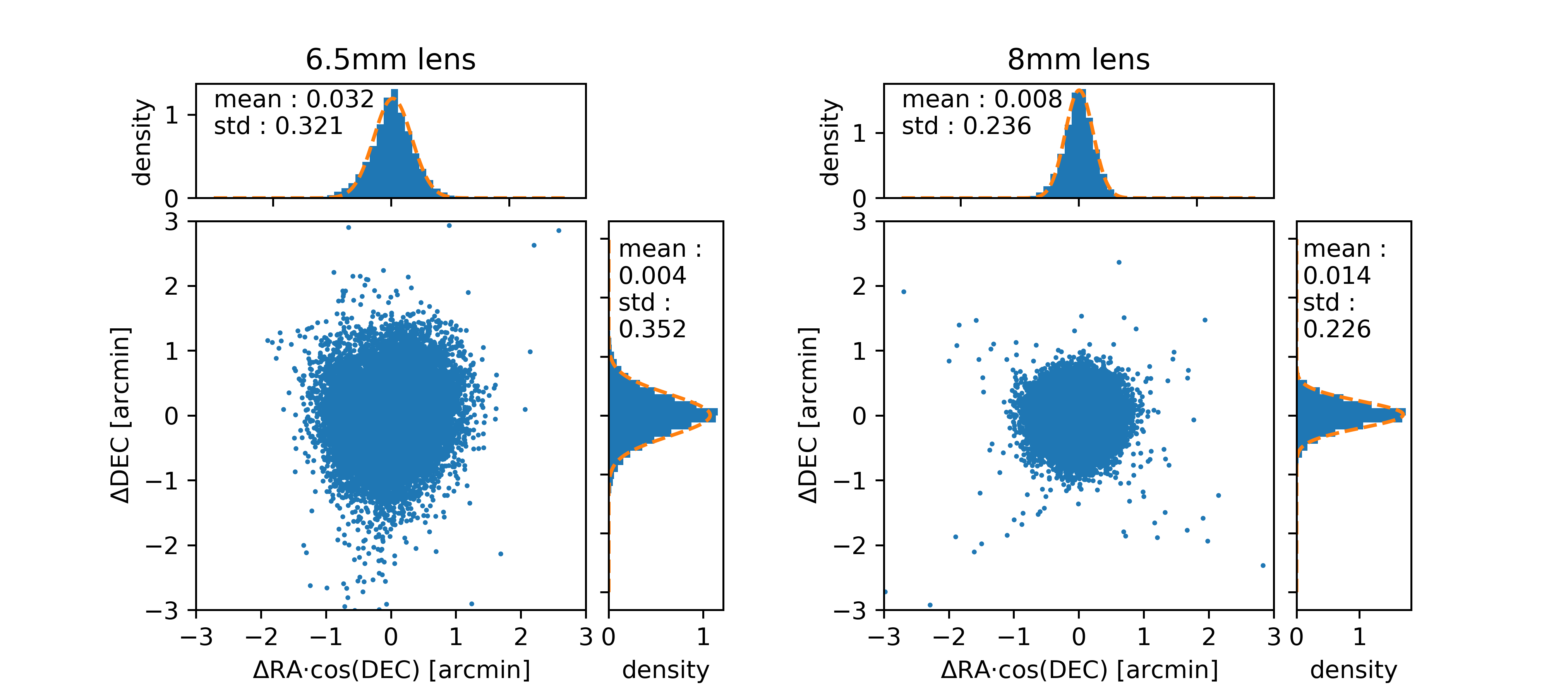}
    \caption{Distribution of astrometric fitting residues of 6.5mm (left) and 8mm (right) lens. The Gaussian fitting along R.A. and decl. direction are shown in the subplots as orange dashed lines.
    }
    \label{fig:astrometry}
\end{figure}

\begin{figure}[!htb]
    \centering
    \includegraphics[width=0.9\textwidth]{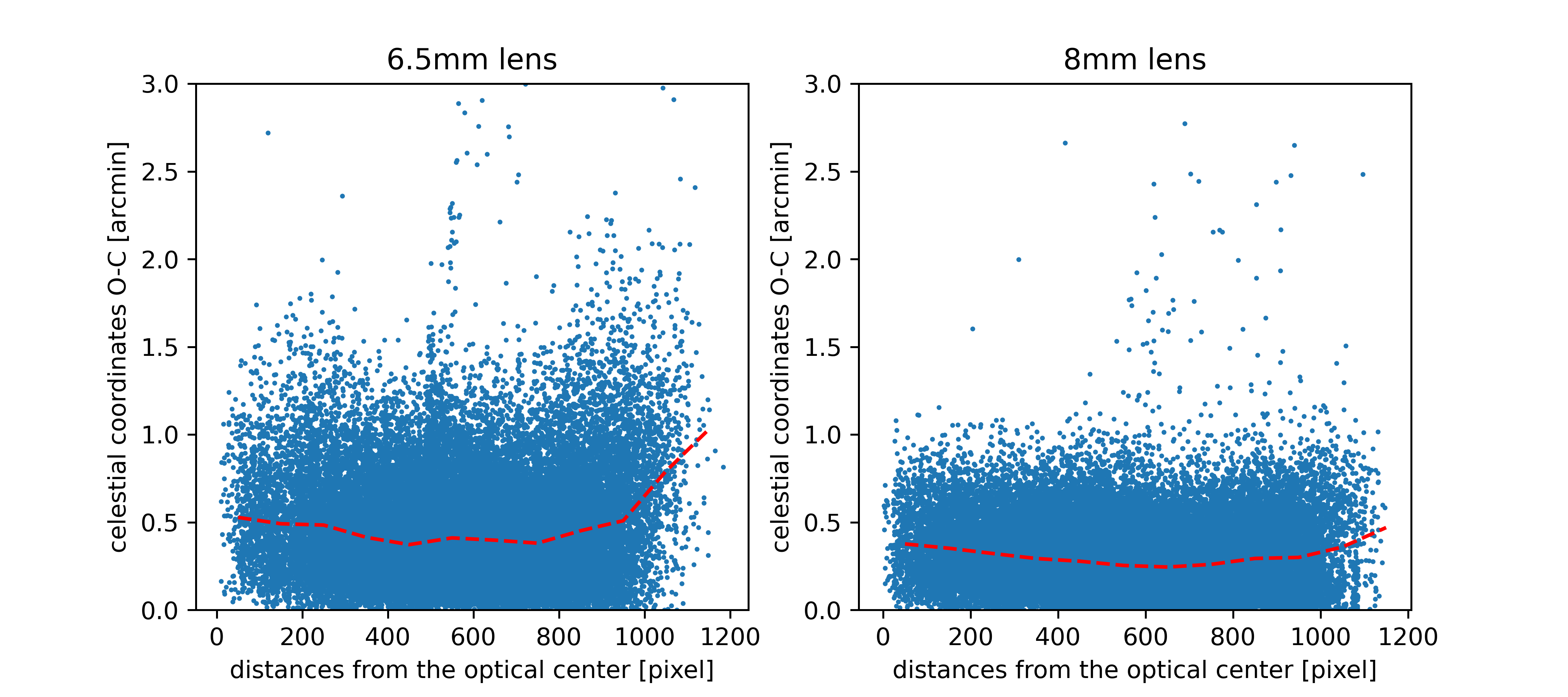}
    \caption{Distribution of astrometric fitting residues of 6.5mm (left) and 8mm (right) lens, with respect to the distances of the stars from the image center. The medians of the bins for every 100 pixels are shown in red dashed lines.
    }
    \label{fig:astrometry_center}
\end{figure}

\begin{figure}[!htb]
    \centering
    \includegraphics[width=0.9\textwidth]{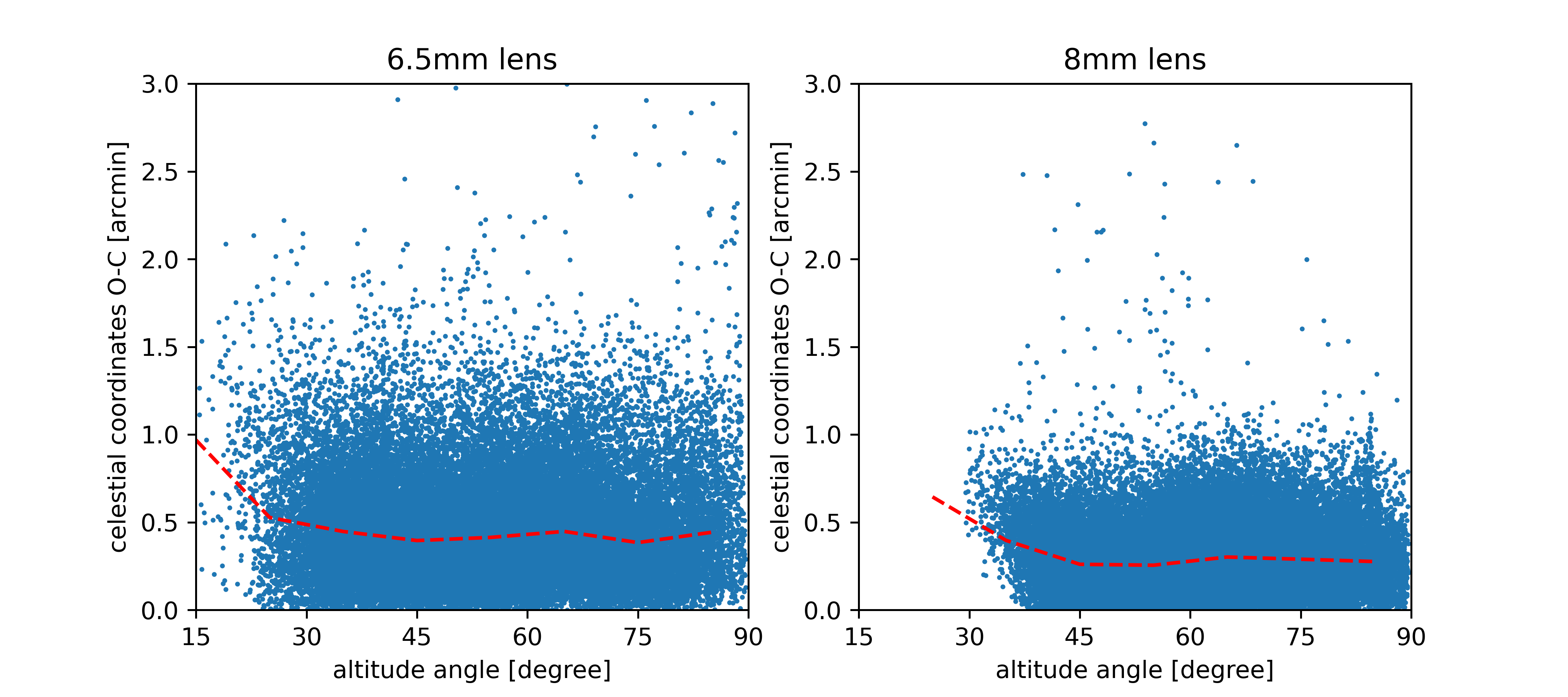}
    \caption{Distribution of astrometric fitting residues of 6.5mm (left) and 8mm (right) lens, with respect to the altitudes of the stars. The medians of the bins for every 15 degrees are shown in red dashed lines.
    }
    \label{fig:astrometry_altitude}
\end{figure}

The medians of fitting residues of the stars are 0.42 arc-minutes for the 6.5mm lens and 0.28 arc-minutes for the 8mm lens, respectively, and their distributions are shown in Fig. \ref{fig:astrometry}. In this figure, our results show that 8mm lenses have smaller standard deviations on both RA and DEC directions. This shows that 8mm lenses are better than 6.5mm lenses for astrometric precision. The distribution of astrometric fitting residuals with respect to the distances of the stars from the image center and their altitude angles is illustrated in Figures \ref{fig:astrometry_center} and \ref{fig:astrometry_altitude}. Fig. \ref{fig:astrometry_center} indicates that for 6.5mm lenses, the fitting residuals begin to increase when the distance exceeds 1,000 pixels, suggesting reduced accuracy in the extreme corners of the image. In contrast, for 8mm lenses, the fitting accuracy at the corners of the image is comparable to that in other areas. Additionally, Fig. \ref{fig:astrometry_altitude} shows that the residuals are higher for stars at lower altitudes. Given the orientation of the fields of view during the test, these low-altitude stars are also positioned near the corners of the images.


Additionally, the limiting magnitude of the cameras is compared. During the test observation, 100 frames with 33.3ms exposure time were captured and stacked as the test sample. The gain setting of the cameras is set to 361 which represents 2.2 e$^-$/ADU. In the stacked frames, pixel positions of the stars were projected according to the coordinates in the HIPPARCOS Catalogue \citep{1997A&A...323L..49P}. Then the magnitudes and SNRs (signal-to-noise ratios) for these stars were obtained using aperture photometry. The limiting magnitude is defined by the magnitude when half of the stars have SNRs greater than 3. The limiting magnitude is 4.07 for the 6.5mm lens, and 5.58 for the 8mm lens, respectively. This shows a significant increase in the capability of detecting meteors using the 8mm lens, and explained that even with a narrower FOV, the 8mm lens can detect more meteors than the 6.5mm lens.

\subsection{System Performance}

The four stations in Beijing entered regular operations after being installed. The observation is highly automated, and the stations have been running continuously for a year. The only short interruptions happened during the power and network loss and the computer system updates. In early August 2023, severe flooding interrupted operations at the Baihuashan station, resulting in a loss of network connection for more than a month.

\begin{figure}[!htb]
    \centering
    \includegraphics[width=0.6\textwidth]{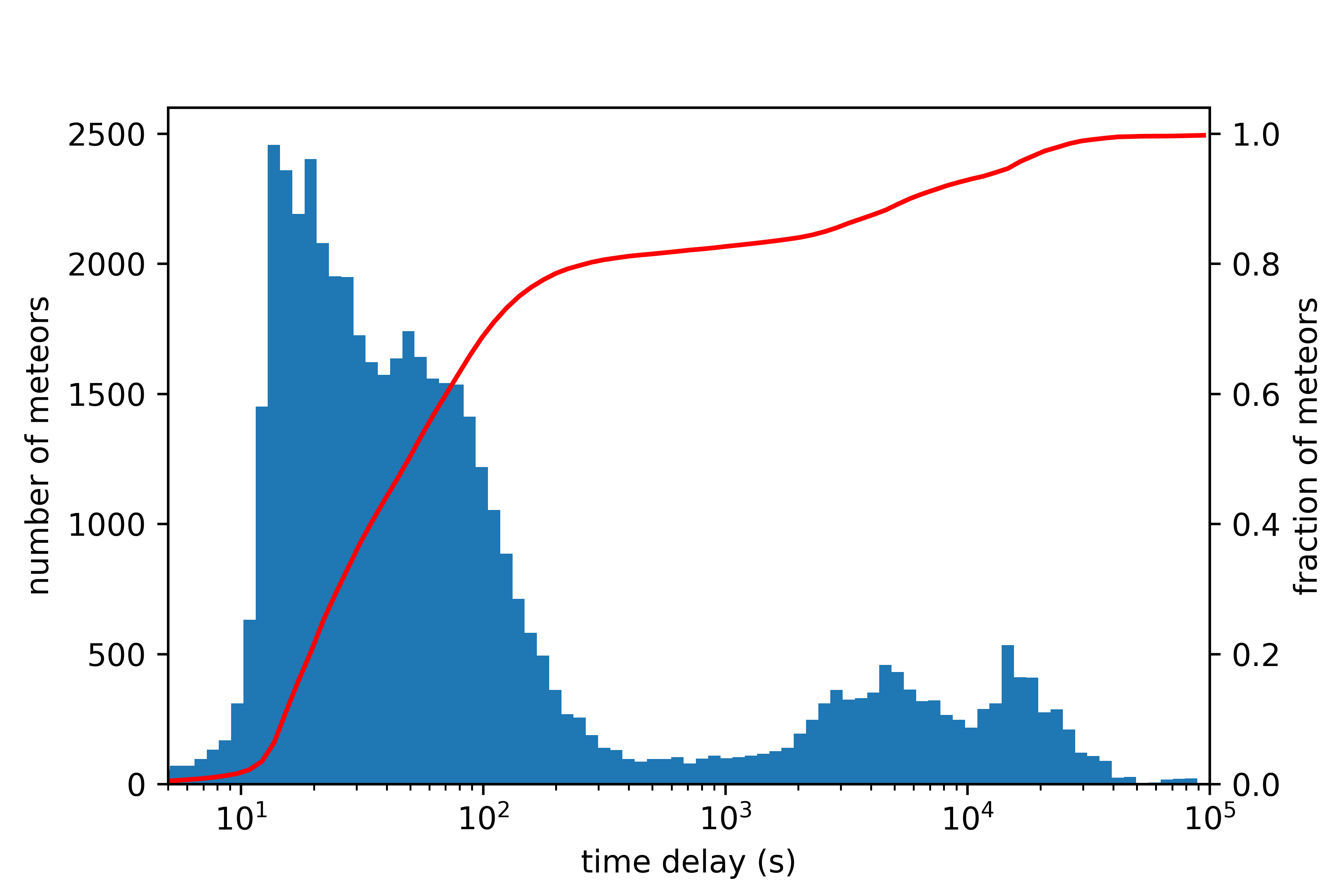}
    \caption{The distribution of processing time. After the meteors are detected, 70\% of the meteors are processed within 100 seconds. The remaining 30\% of them are finished within 10 hours.
    }
    \label{fig:ptime}
\end{figure}

The observed data generated by \textit{meteorThread} were processed by \textit{meteorExtractor} directly in computer memory, and this can accelerate the process of the observation. If the queue was not congested, the coordinates of most meteors can be obtained in $\sim$100 seconds. The queue is often congested during twilight when many satellites are also observed, which increase the computational demand. In this case, the meteor data are cached in storage and gradually processed with a delay of hours. The distribution of processing duration is shown in Fig. \ref{fig:ptime}. From this figure, we can see that, after the meteors are detected, 70\% of the meteors are processed within 150 seconds, and the remaining 30\% are finished within 10 hours. On the data platform, it takes $\sim$10s on average to calculate the orbit of a meteoroid.

\begin{figure}[!htb]
    \centering
    \includegraphics[width=0.9\textwidth]{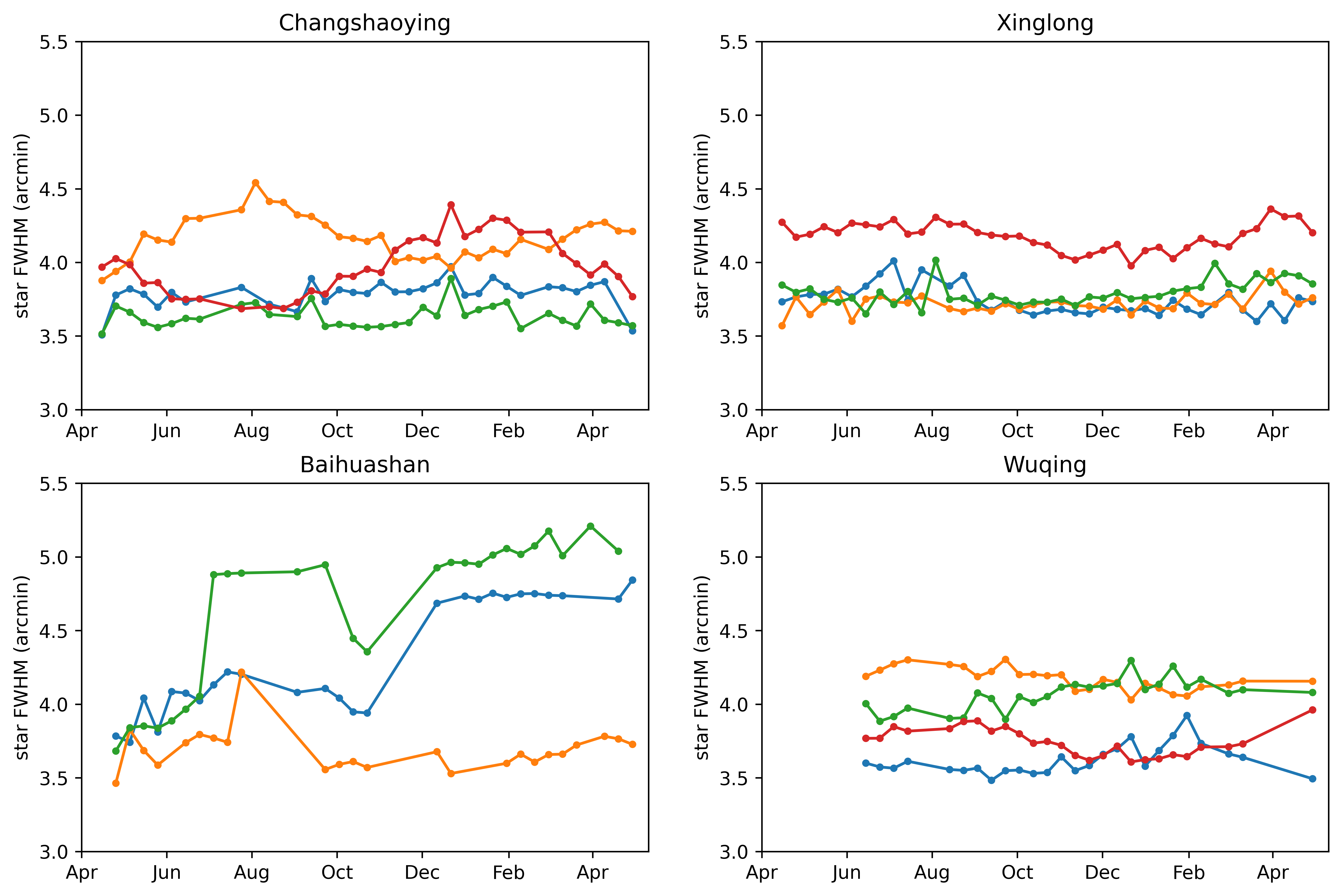}
    \caption{The change of the FWHM for each camera over one year of operation.}
    \label{fig:fwhm}
\end{figure}

The optical performance of the cameras is monitored through measurements of the stars in the images. The FWHM of the sources mainly depends on the lens focusing since for wide angle cameras the FWHM is much larger than the astronomical seeing. Thus the FWHM may change over extended periods. The change of the FWHM for each camera is shown in Fig. \ref{fig:fwhm}. In this figure, the FWHM of the cameras shows slight changes over several months with the exception of the two cameras at Baihuashan station. The focusing of the two cameras degraded, resulting in a significant increase in the FWHM, as shown in lower-left panel of Fig. \ref{fig:fwhm}

\begin{figure}[!htb]
    \centering
    \includegraphics[width=0.9\textwidth]{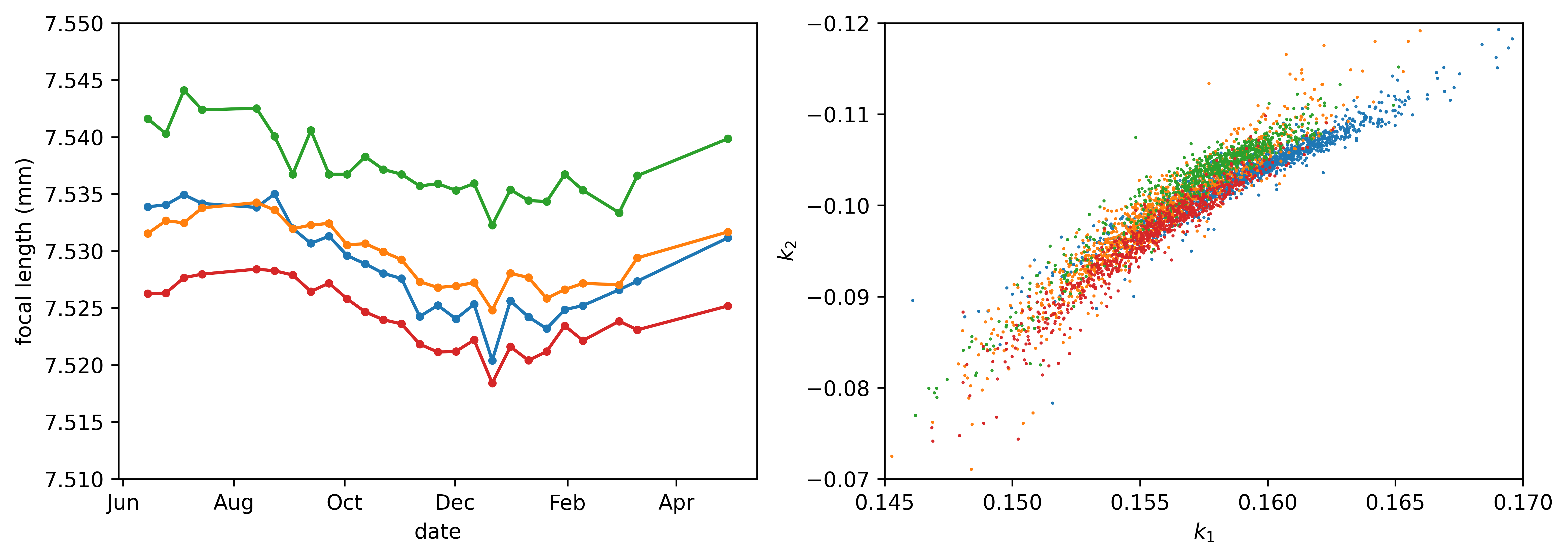}
    \caption{Left: The change of the focal lengths for each camera at Wuqing Station. Right: The distribution of fitted $k_1$ and $k_2$ coefficients for each camera.}
    \label{fig:camera_model}
\end{figure}

Fig. \ref{fig:camera_model} shows the differences in parameters among the cameras at one of the stations as an example. Although the focal lengths vary for each camera, they change uniformly in response to air temperatures. The distribution of the two polynomial coefficients differs across all cameras and does not change systematically over time.

\subsection{Meteor trajectories}

\begin{figure}[!htb]
    \centering
    \includegraphics[width=1\textwidth]{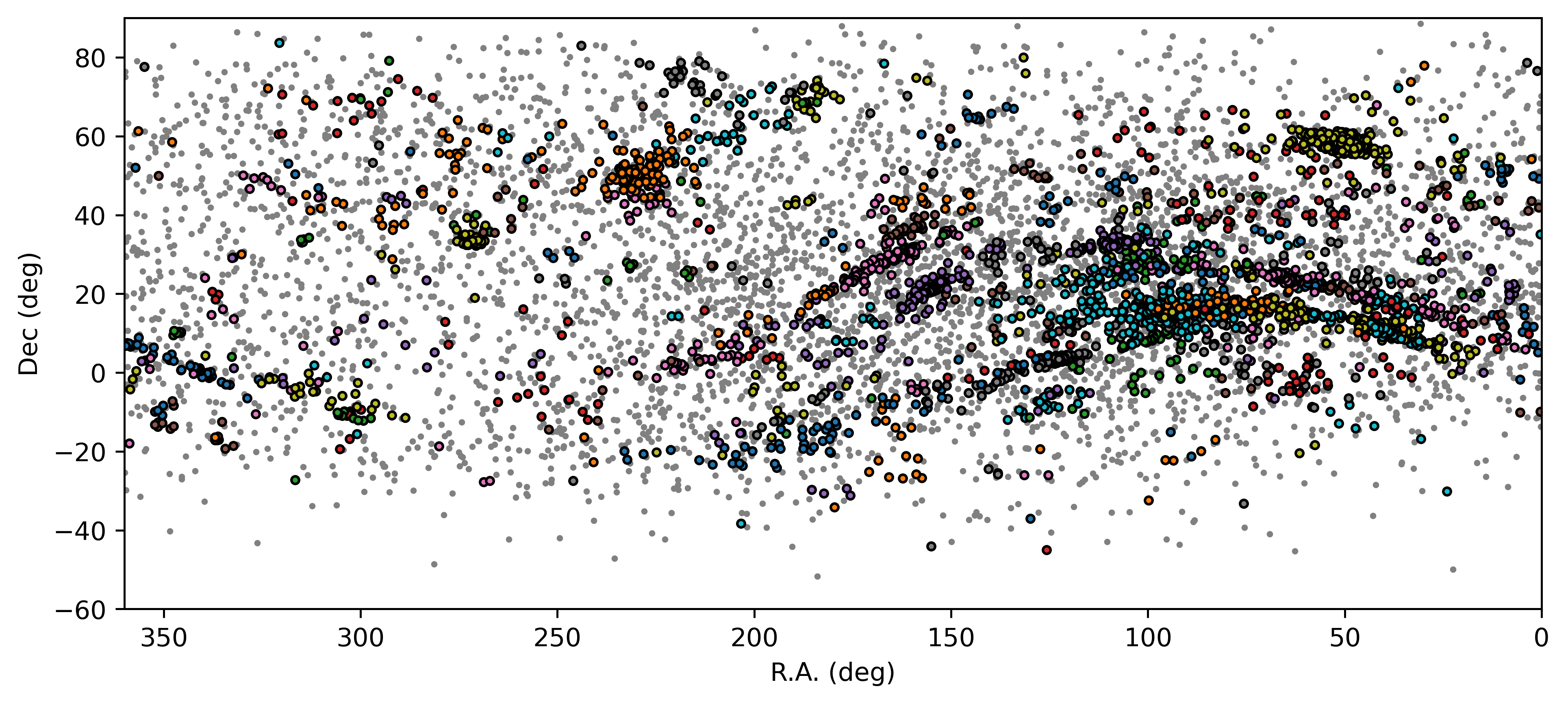}
    \caption{The radiants of all the meteors detected by the network in this paper. Meteors belong to known meteoroid streams are circled by black. 
    }
    \label{fig:radiants}
\end{figure}

\begin{figure}[!htb]
    \centering
    \includegraphics[width=0.5\textwidth,trim=3cm 2cm 3cm 1.5cm,clip]{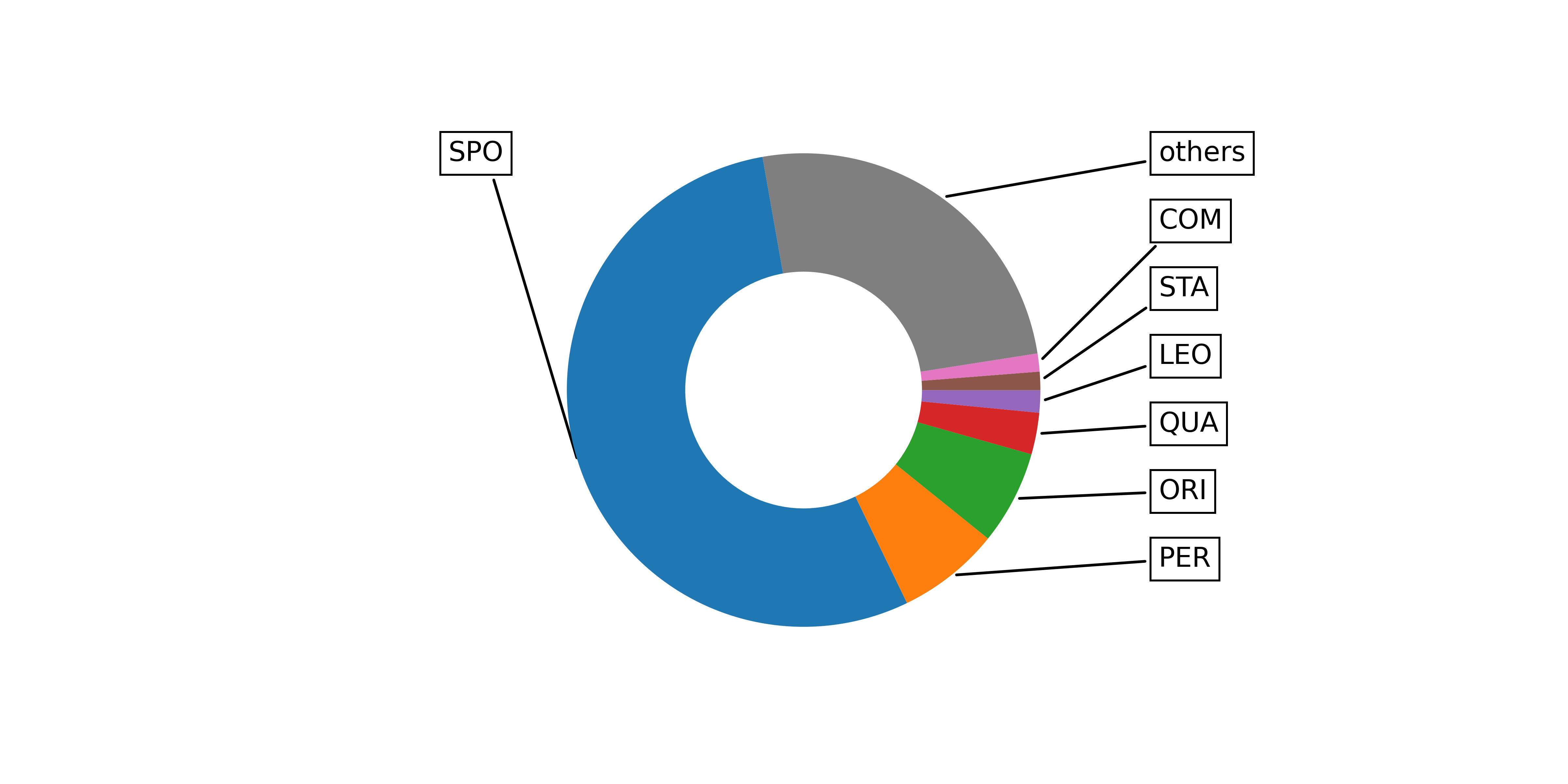}
    \caption{The composition of detected meteors based on meteoroid streams. Nearly half of the meteors are sporadic, and most of the meteors from known meteoroid streams are Perseid, Orionid, and Quadrantid.
    }
    \label{fig:groups}
\end{figure}

During the operation of the network of 4 stations around Beijing, 8,683 orbits of the meteoroids were successfully obtained during 10 month period. 4,725 of them can be linked to 265 established meteoroid streams, as recorded by IAU Meteor Data Center\footnote{\url{https://www.ta3.sk/IAUC22DB/MDC2022/Roje/roje_lista.php?corobic_roje=0&sort_roje=0}} \citep{2023A&A...671A.155H}. The radiants of all the meteors are shown in Fig. \ref{fig:radiants}. In this figure, the clustering of the meteor from known meteoroid streams is clearly visible, as they are plotted in circled and colored dots. The distribution of radiants of sporadic meteors is relatively uniform. The distribution of affiliating meteoroid streams of meteors is shown in Fig. \ref{fig:groups}. This figure shows clearly that, nearly half of the meteors are sporadic (SPO) and most of the meteors from known meteoroid streams are Perseid, Orionid, and Quadrantid.


\begin{figure}[!htb]
    \centering
    \includegraphics[width=0.9\textwidth]{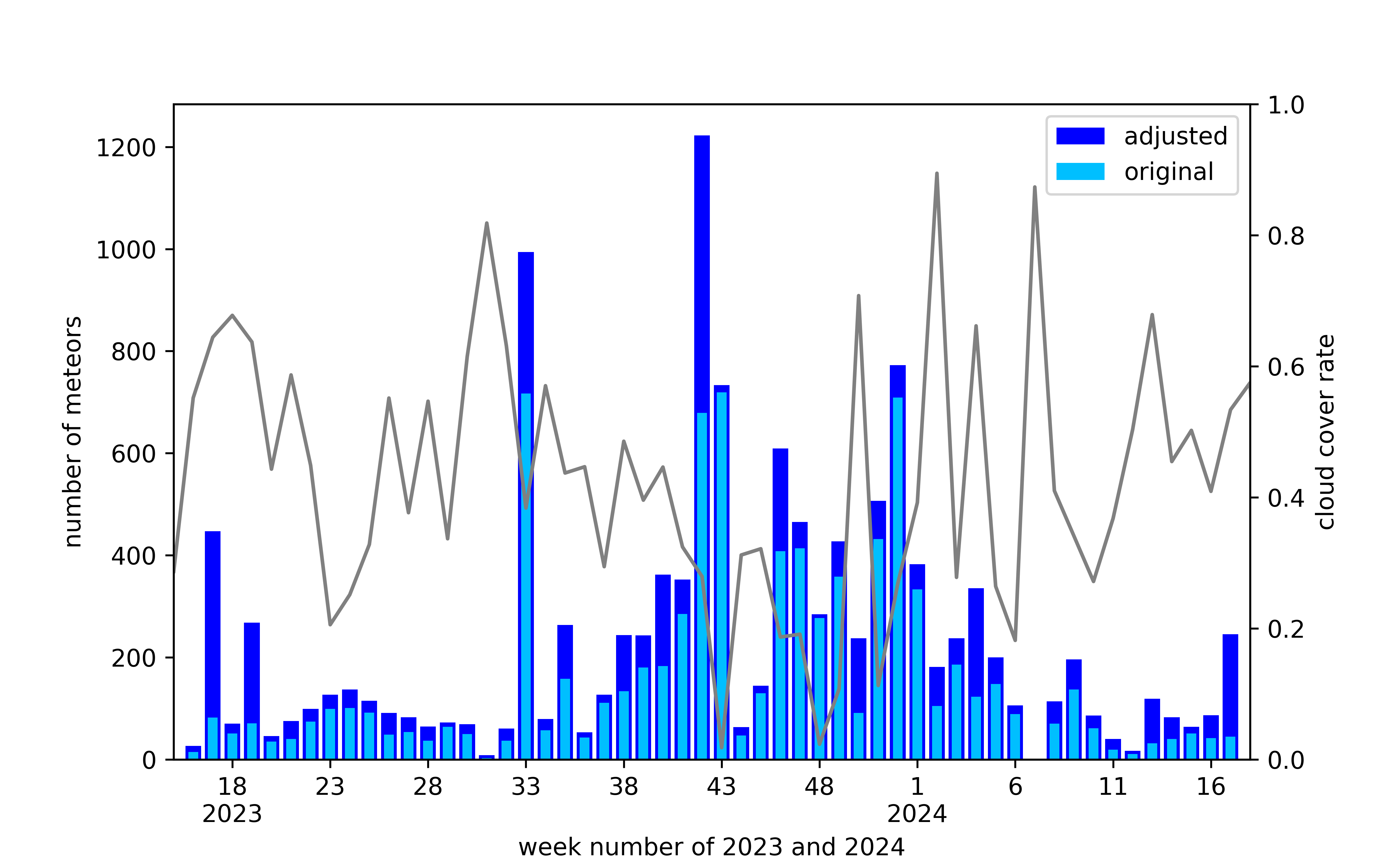}
    \caption{Number of multi-station} meteor detected in every week (light blue) and the value adjusted by cloud cover rate (blue). The cloud cover rate is plotted in a gray line. Three spikes at the week 33, 42 and 1 represent three major meteor showers observed, which are Perseid, Orionid, and Quadrantid.
    \label{fig:byweek}
\end{figure}

The detection rate of meteors is heavily influenced by the weather. \cite{2020A&A...644A..53C} used cloud cover rates to adjust meteor detection number to get a better representation of natural meteor rates. We did the similar adjustment to the weekly accumulated numbers of meteors detected by the network using the hourly cloud coverage of reanalysis product provided by the Copernicus Climate Change Service \citep[C3S\footnote{\url{https://cds.climate.copernicus.eu/datasets/reanalysis-era5-single-levels?tab=overview}};][]{ecmwf}. Fig. \ref{fig:byweek} shows the weekly accumulated numbers of multi-station meteors detected and the corresponding value adjusted by the cloud cover rates. The weather is clear during the Perseid, Orionid, and Quadrantid meteor showers, so the sharp spikes are observed. However, due to the bad weather in mid-December 2023, the Geminid meteor shower in 2023 was not well observed. 

\begin{figure}[!htb]
    \centering
    \includegraphics[width=1.0\textwidth]{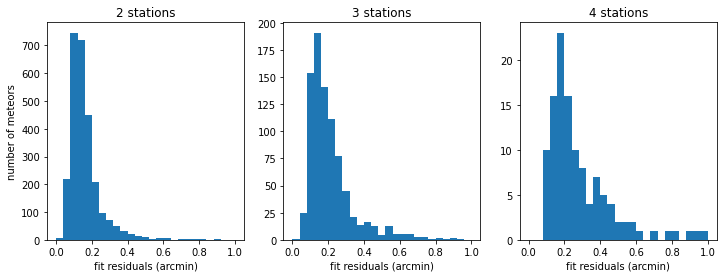}
    \caption{The distribution of the angular residuals from the fitted positions categorized by the number of observing stations.}
    \label{fig:angle_residual}
\end{figure}

The angular deviations of the observed positions from the fitted positions are calculated for each sporadic meteor trajectory to determine the uncertainties. Fig. \ref{fig:angle_residual} illustrates the distribution of these angular residuals, categorized by the number of observing stations. The median residual is 0.142 arc-min. Specifically, the median residuals for meteors observed by 2, 3, and 4 stations are 0.132, 0.175, and 0.218 arc-min, respectively. The increase in deviation with the number of stations suggests the presence of systematic fitting errors, which is also reported by \cite{2021GMN}. Notably, the angular deviation remains comparable to that of high-precision 16mm lenses used by GMN \citep{2021GMN}, despite having a significantly larger field of view ($77^\circ\times51^\circ$ vs. $20^\circ\times11^\circ$).

\begin{figure}[!htb]
    \centering
    \includegraphics[width=0.6\textwidth]{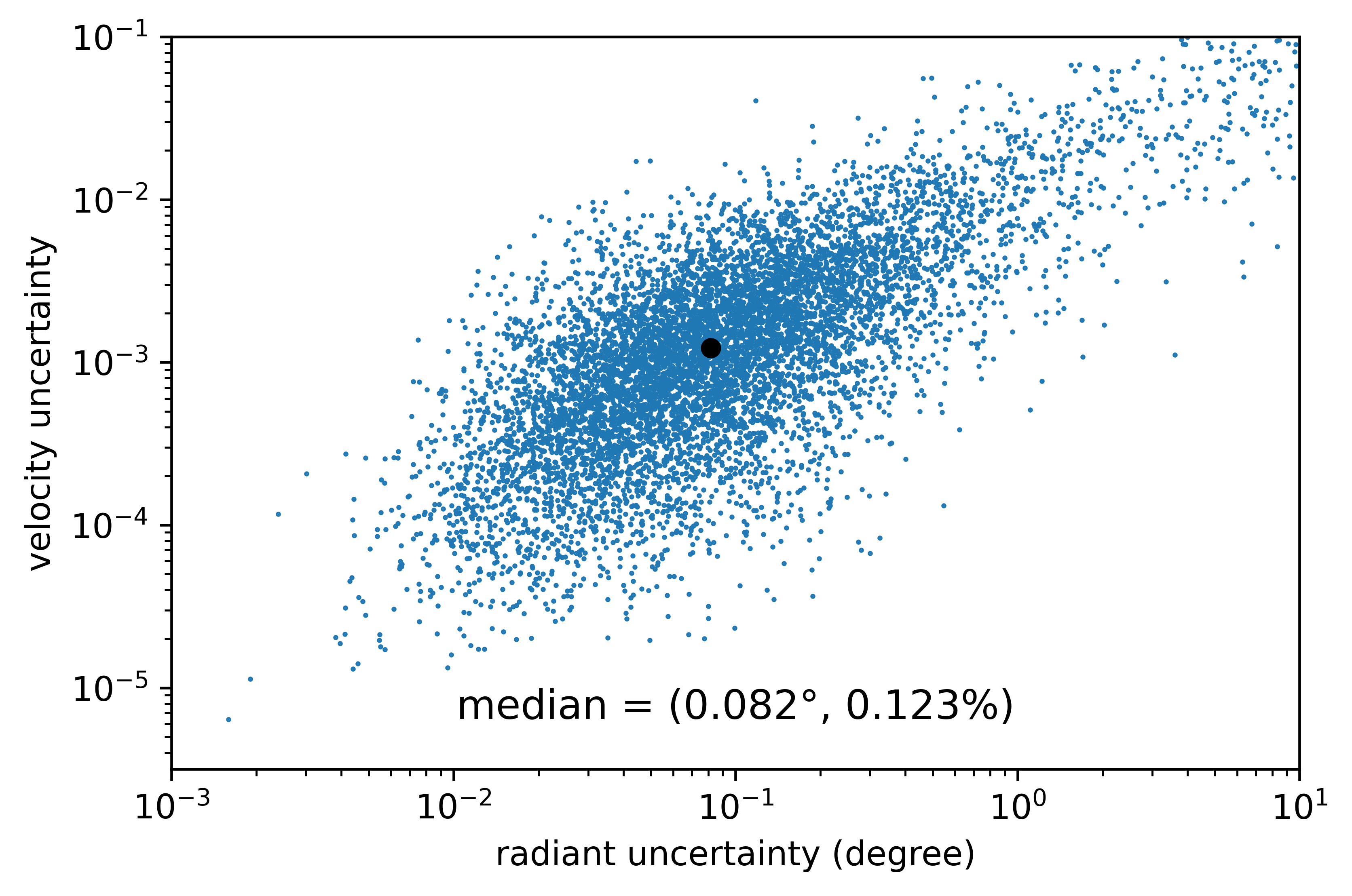}
    \caption{The distribution of the estimated uncertainties of the radiants and velocities}
    \label{fig:uncertainties}
\end{figure}

The Monte Carlo method is utilized to determine meteor trajectories, aiming to improve the representation of meteoroid dynamics \citep{2020MNRAS.491.2688V}. Through this process, the uncertainties of the radiants can be estimated. Figure \ref{fig:uncertainties} illustrates the distribution of position uncertainties for the radiants and the relative uncertainties of the velocities, derived from 50 Monte Carlo runs. The median uncertainties are 0.082$^\circ$ for position and 0.123\% for velocity, which represent a significant improvement over the 16mm lens used by GMN \citep{2021GMN}.

These improvements in determining meteor trajectories can be attributed to the use of USB cameras that generate raw image data. As demonstrated in the previous study \citep{2024li}, the fitting residuals of meteoroid trajectories are doubled when compressed using H.264 encoding. The introduction of additional noise during processing and compression by IP cameras can ultimately increase the uncertainties of the radiants.


\subsection{Photometry}
We developed a software component for \textit{meteorExtractor} to perform photometry for the detected meteors. The photometry of the meteors uses stars as references. The flux of the stars is obtained through aperture photometry, where the background flux is estimated by circular annuli around the stars. The flux of the meteors is measured by apertures derived from the shape of the meteor streak, and the background flux is estimated by the average of frames without the meteors. This process is done by using the Python library \textit{Photutils} by \cite{larry_bradley_2023_7946442}.

In order to get the $V$-band magnitudes from the instrument magnitudes of the meteors, we use the instrument magnitudes of the stars and $V$-band magnitudes from the HIPPARCOS Catalogue \citep{1997A&A...323L..49P}. Two key parameters are considered: air extinction coefficient, and color. Within the wide field image, air extinction is not the same for all positions, and it is heavily influenced by the altitude angle of the sources. In additions, the camera covers a larger wavelength range than the normal $V$-band, so a correction which is proportional to $V-I$ color index is also required. The photometry of the meteors can be described by the equations below:

\begin{equation}
M_V = M_{\rm{ins}} + r \cdot airmass + k \cdot C_{V-I} + C,
\end{equation}

\begin{equation}
     M_{\rm{ins}}  = -2.5 \rm{log} \it{f}, 
\end{equation}
where $f$ is the measured fluxes of sources from the image, $r$ is the air extinction coefficient, $k$ is the coefficient for the color index and $C$ is the zero point. During the measurement of the positions of the meteors, the fluxes of meteors and stars can be obtained at the same time. The airmass of stars and meteors can be calculated from the altitude angle using the equations in \cite{1962aste.book.....H}, which can be calculated by their coordinates measured from the image. The color of stars are provided by the star catalog. Then, the parameters $r$, $k$ and $C$ are fitted based on the $V$-band magnitude, airmass, and color of the stars. Finally, the apparent $V$-band magnitude of the meteor in each frame is obtained. The parameters $r$, $k$, and $C$ are fitted separately for each detected meteor, as air extinction varies with weather conditions.

\begin{figure}[!htb]
    \centering
    \includegraphics[width=0.6\textwidth]{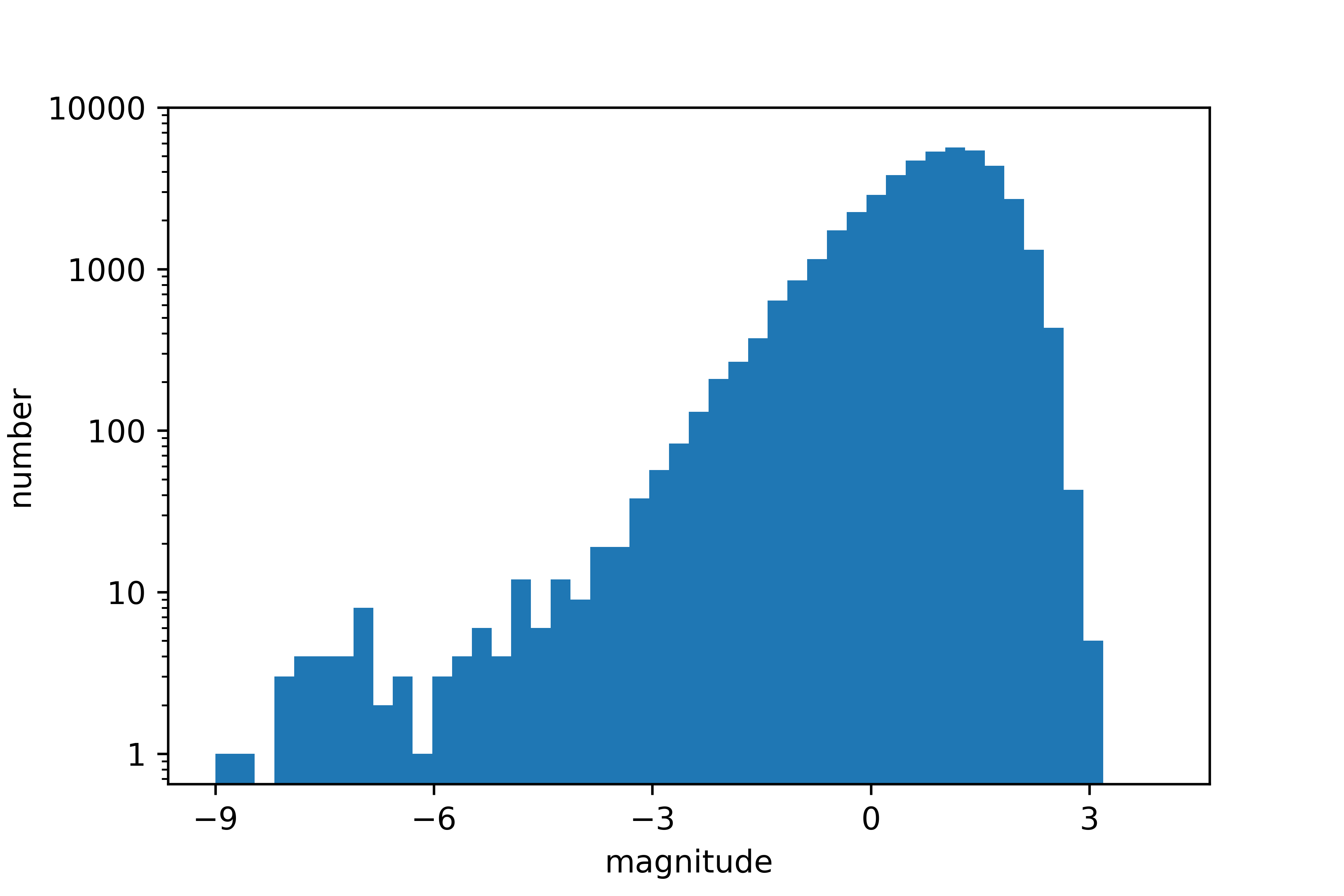}
    \caption{The distribution of $V$-band magnitudes of the meteors.
    }
    \label{fig:phot}
\end{figure}

The distribution of the apparent magnitudes of the detected meteors are shown in Fig. \ref{fig:phot}. In this figure, the magnitude of meteors is represented by the average value of its light curve. Fig. \ref{fig:phot} shows that the numbers of meteors rise with the increase of magnitude before reaching the peak at $\sim$1 mag, which corresponds with the distribution of the size of meteoroids. And the peaks at $\sim$1 mag indicates the detection limit of meteors of our system, which is lower than the limiting magnitude of stars (5.58 mag). This is mainly because meteors are not point sources, and the flux is dispersed among more pixels than stars, which introduces more noises.

\begin{figure}[!htb]
    \centering
    \includegraphics[width=0.6\textwidth]{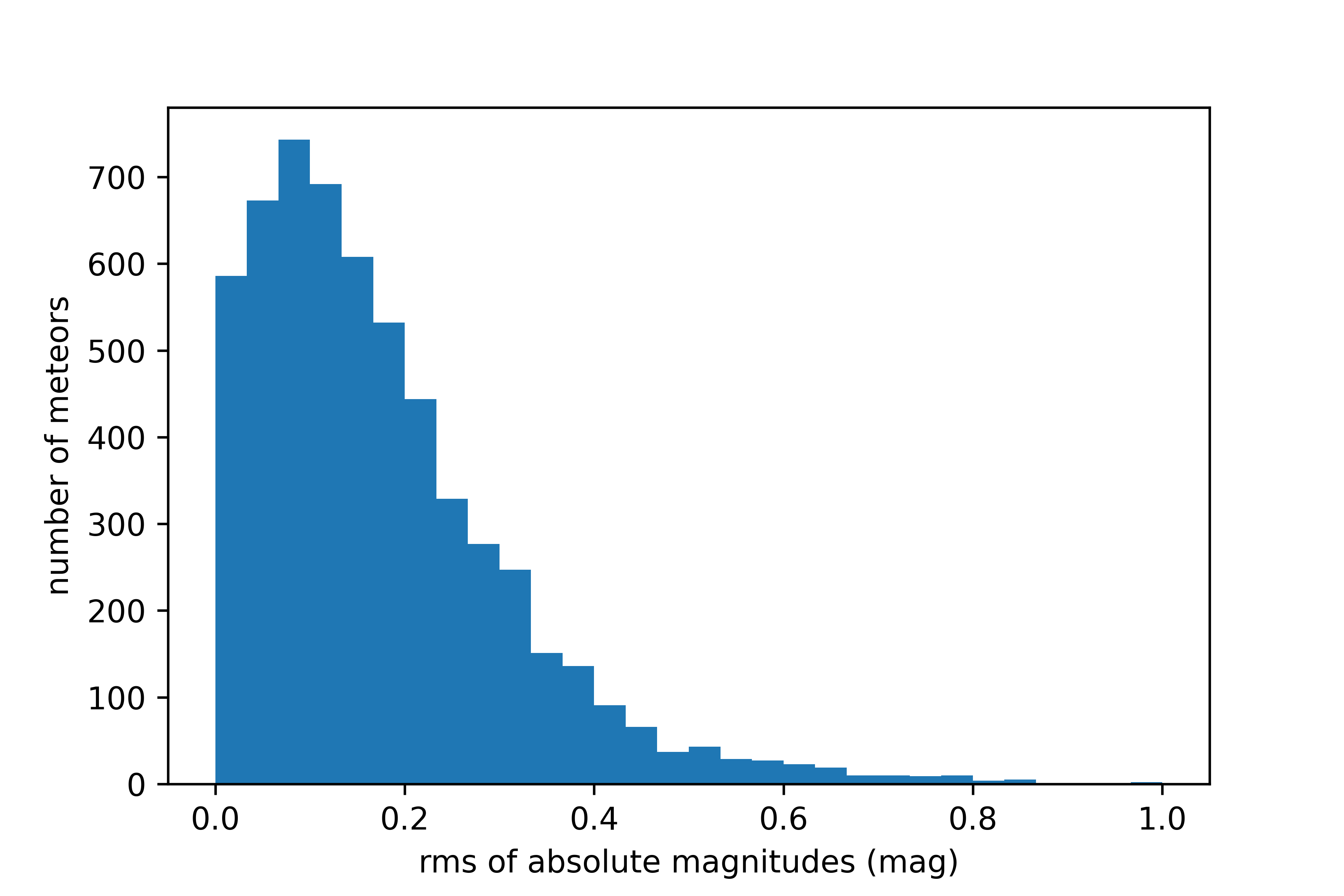}
    \caption{The distribution of RMS of the observed absolute magnitudes of multi-station meteors.}
    \label{fig:phot_rms}
\end{figure}

The uncertainties of photometric measurements can be estimated using multi-station meteors. The absolute magnitudes are calculated separately using the distances from each station. The distances from the stations are significantly more accurate and do not introduce measurable uncertainties to the absolute magnitudes. Therefore, the RMS values of the absolute magnitudes from each station can be used to estimate the uncertainties of the photometric results. Figure \ref{fig:phot_rms} illustrates the distribution of the RMS values for the observed absolute magnitudes of multi-station meteors, with a median value of 0.144 mag.

\subsection{Sporadic Meteoroids}

\begin{figure}[!htb]
    \centering
    \includegraphics[width=1\textwidth]{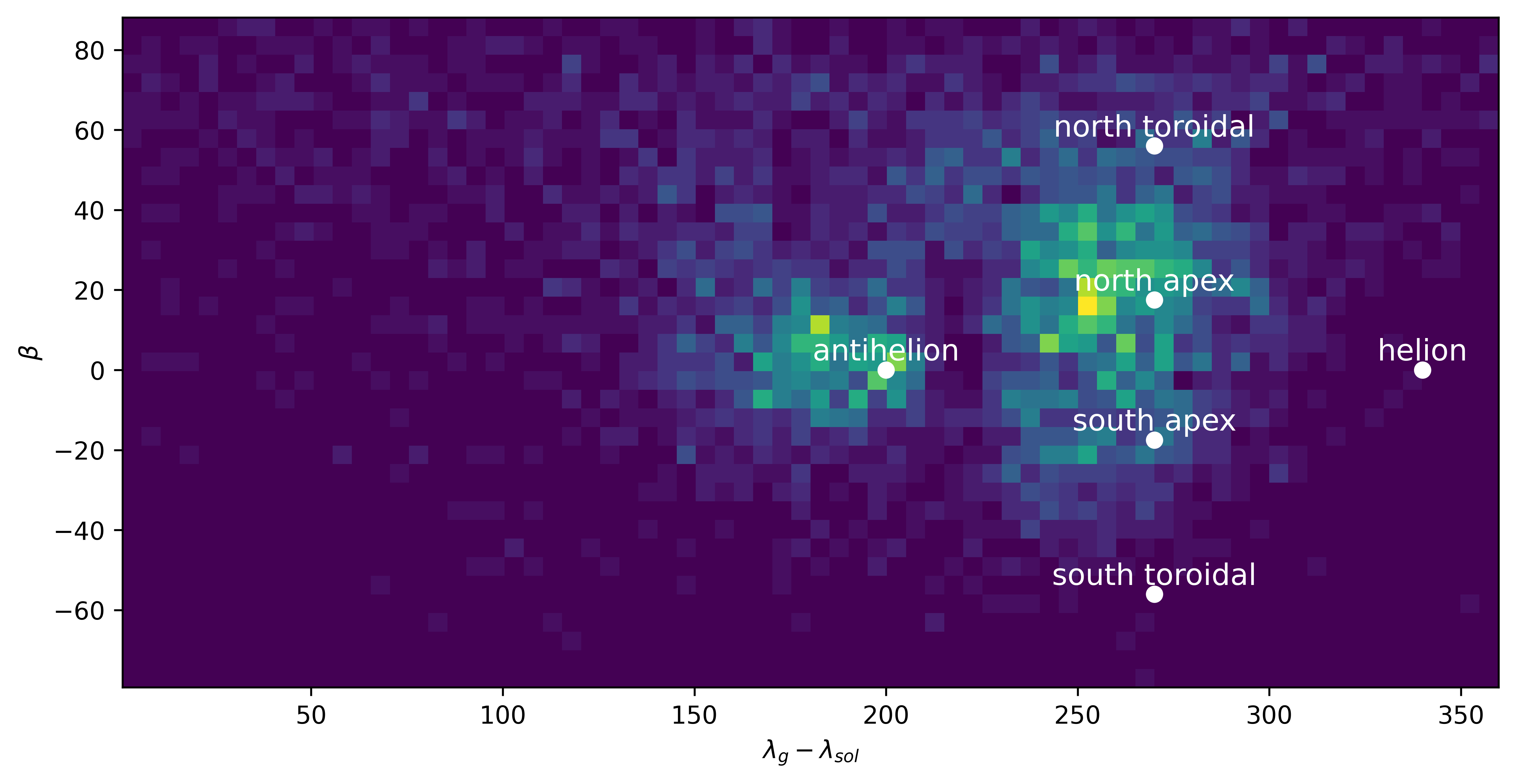}
    \caption{The distributions of radiants of sporadic meteoroids in a Sun-centered ecliptic frame. White dots are observed sources by \cite{2008Icar..196..144C} as reference.
    }
    \label{fig:spo_source}
\end{figure}

Meteoroid engineering models are developed to describe the risk posed by meteoroids to spacecrafts \citep{2019LPICo2109.6054M}. In the models, sporadic meteoroids are the main concern as the risk of collisions accumulates over long time span. In the NASA MEM3 \citep{2019LPICo2109.6054M}, sporadic meteoroids are described by three populations: helion and antihelion, apex, and toroidal \citep{2008Icar..196..144C}. In our results involving 3,958 sporadic meteoroids, most of the populations except helion ones are visible, as shown in Fig. \ref{fig:spo_source}. In Fig. \ref{fig:spo_source}, the relative positions of the radiants and the Sun in ecliptic coordinates are plotted, as the horizontal axis being the difference of ecliptic longitude of the radiants and the Sun ($\lambda_{\rm{g}}-\lambda_{\rm{sol}}$) at the time of the observations, and the vertical axis being the elliptic latitude ($\beta$). The lack of helion meteoroids can be explained by the fact that optical observations only occur at night and away from the Sun.

\begin{figure}[!htb]
    \centering
    \includegraphics[width=1\textwidth]{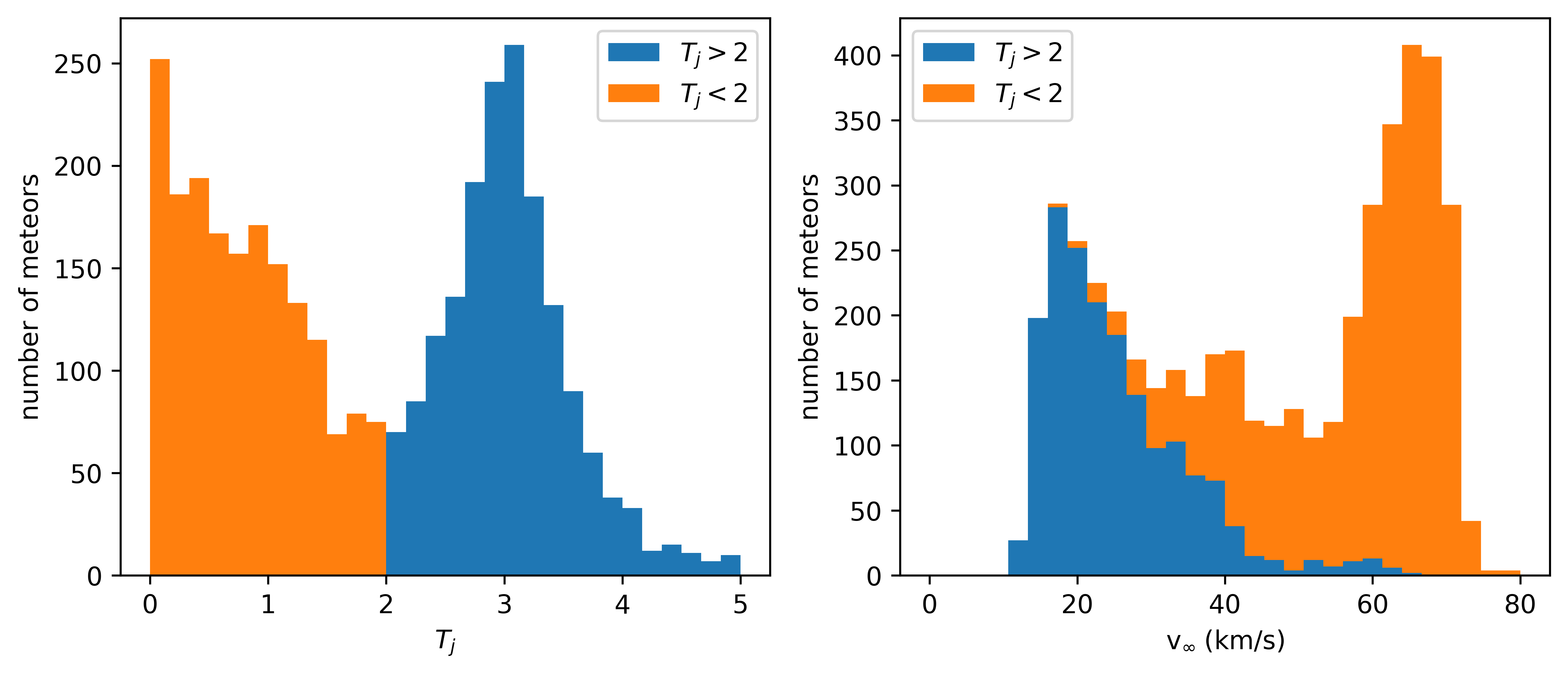}
    \caption{The distribution of the $T_j$ and speed of sporadic meteoroids. The populations of $T_j > 2$ and $T_j <2$ are colored differently in the speed distribution. The correspondence of the two populations and their speed are clearly visible in this figure.
    }
    \label{fig:spo_speed}
\end{figure}

\cite{2019LPICo2109.6054M} also used Tisserand parameters, $T_j$, to classify the meteoroids by their orbits as asteroidal/ecliptic or cometary/isotropic. The distribution of $T_j$ and speed of the observed meteoroids here is shown in Fig. \ref{fig:spo_speed}. From Fig. \ref{fig:spo_speed} we can see two peaks at $T_j$=0 and 3, indicating the different orbits of parent bodies (isotropic and ecliptic), which are divided by $T_j$=2. Two peaks are also visible in the speed distribution shown in Fig. \ref{fig:spo_speed} at $\sim$20 km/s and $\sim$65 km/s which is similar to those obtained by \cite{2020A&A...644A..53C}. These two peaks in the speed distribution are clearly corresponding to different $T_j$ populations.

The distribution of densities of meteoroids are derived from the optical observations by modelling their ablation and deceleration \citep{EHLERT2020249,2011A&A...530A.113K,2019LPICo2109.6054M}. \cite{2011A&A...530A.113K} measured the densities and orbit parameters of 107 meteoroids, and \cite{EHLERT2020249} recently refined their densities, and they reported that meteoroids with $T_j$ above or below 2 have different densities, in which the asteroidal meteoroids with $T_j>$2 have high densities and the cometary ones with $T_j<$2 have low densities. In the MEM3, the distribution of densities is fitted by two Gaussian distributions with average densities of 3,792 kg/m$^{3}$ and 857 kg/m$^{3}$, respectively \citep{2019LPICo2109.6054M}.  In the future, we will further analyze the data gathered by the network and produce detailed density measurements.

\subsection{Fireball Analysis}
\begin{figure}[!htb]
    \centering
    \includegraphics[width=0.8\textwidth]{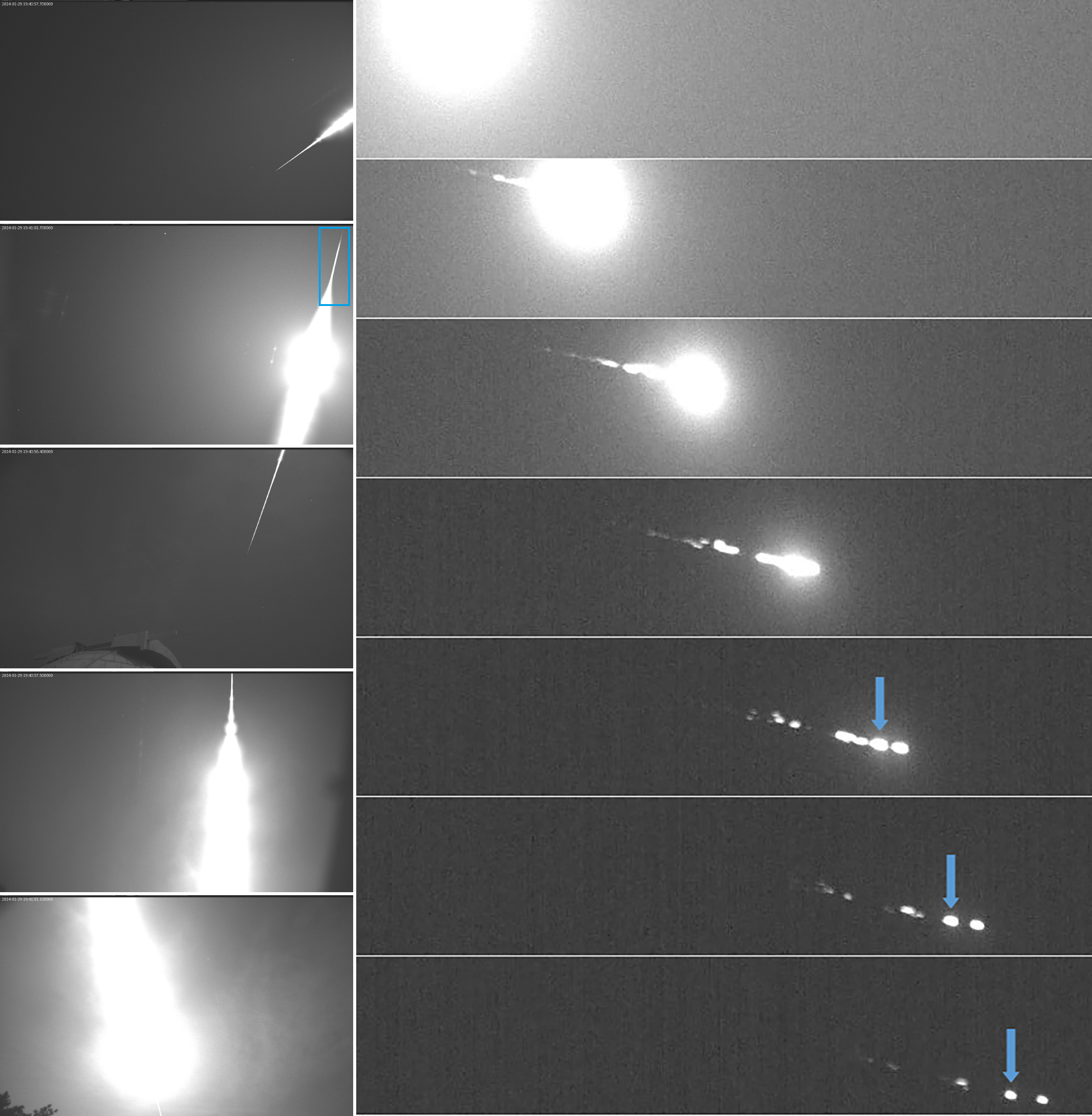}
    \caption{Left: The trajectory of the fireball observed by five cameras at two stations. Right: A sequence of images capturing the fireball after its break-up. The locations of the image patches are indicated by the blue box in the second image of the left column. The images are sampled at 0.3-second intervals. The brightest fragment, used to calculate the landing position, is marked with blue arrows.
    }
    \label{fig:0129image}
\end{figure}

\begin{figure}[!htb]
    \centering
    \includegraphics[width=0.6\textwidth]{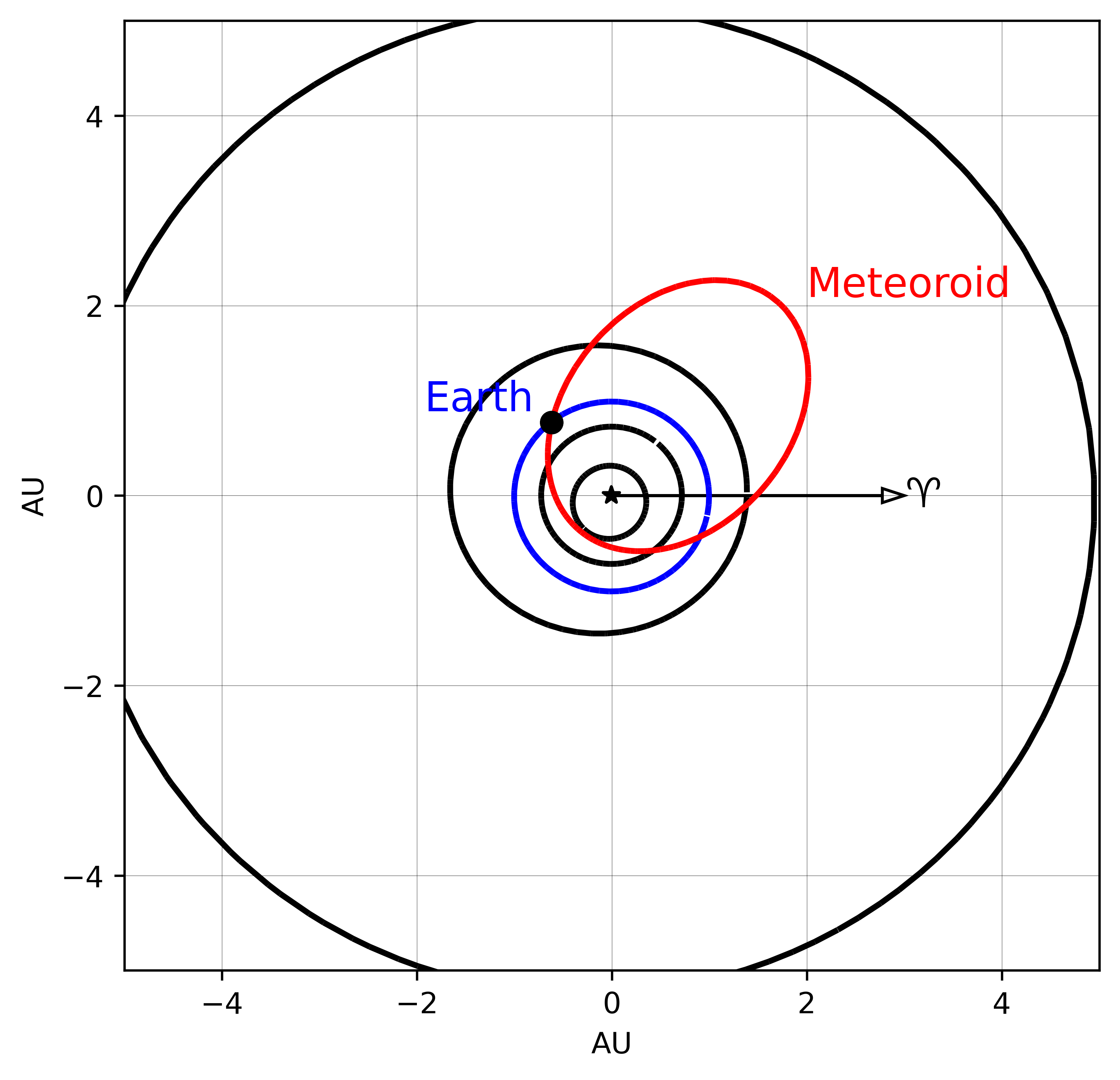}
    \caption{The heliocentric orbit of the observed fireball meteoroids.}
    \label{fig:0129orbit}
\end{figure}

On January 29, 2024, a very bright fireball occurred above the Beijing region and passed within the coverage area of the network. Our network successfully recorded this event, and landing positions of the potential meteorites have been calculated. This fireball was initially detected at 19:40:55 Beijing time, experienced multiple explosions, and disappeared 7.6 seconds later at 19:41:03. Three stations detected this fireball. The image sequence of the fireball post to its break-up is shown in Fig. \ref{fig:0129image}. The trajectory were calculated using the data from the stations. The plot of its orbit is shown in Fig. \ref{fig:0129orbit}, and the corresponding orbital elements are listed in Table \ref{elements}. No meaningful photometric results were obtained due to the over-saturation of the images and the smoggy weather.

\begin{table*}
\centering
\caption{The orbital elements of the fireball meteoroid}
\label{elements}
\begin{tabular}{@{}lll}
\hline
\hline
Orbital elements & Value\\
\hline
Eccentricity ($e$) &  0.686404 +/- 0.001646\\
Semi-major axis ($a$) & 1.591877 +/- 0.008372 AU\\
Inclination ($i$)& 8.288220 +/- 0.015000$^{\circ}$\\
Longitude of the ascending node ($\Omega$) &128.725545 +/- 0.000019$^{\circ}$\\
Argument of perihelion ($\omega$) &102.204811 +/- 0.092645$^{\circ}$\\
Perihelion & 0.499206 +/- 0.000116 AU\\
Aphelion & 2.684549 +/- 0.016798 AU\\
Orbital perioud & 2.008468 +/- 0.015938 years\\
\hline
\end{tabular}
\end{table*}

\begin{figure}[!htb]
    \centering
    \includegraphics[width=0.8\textwidth]{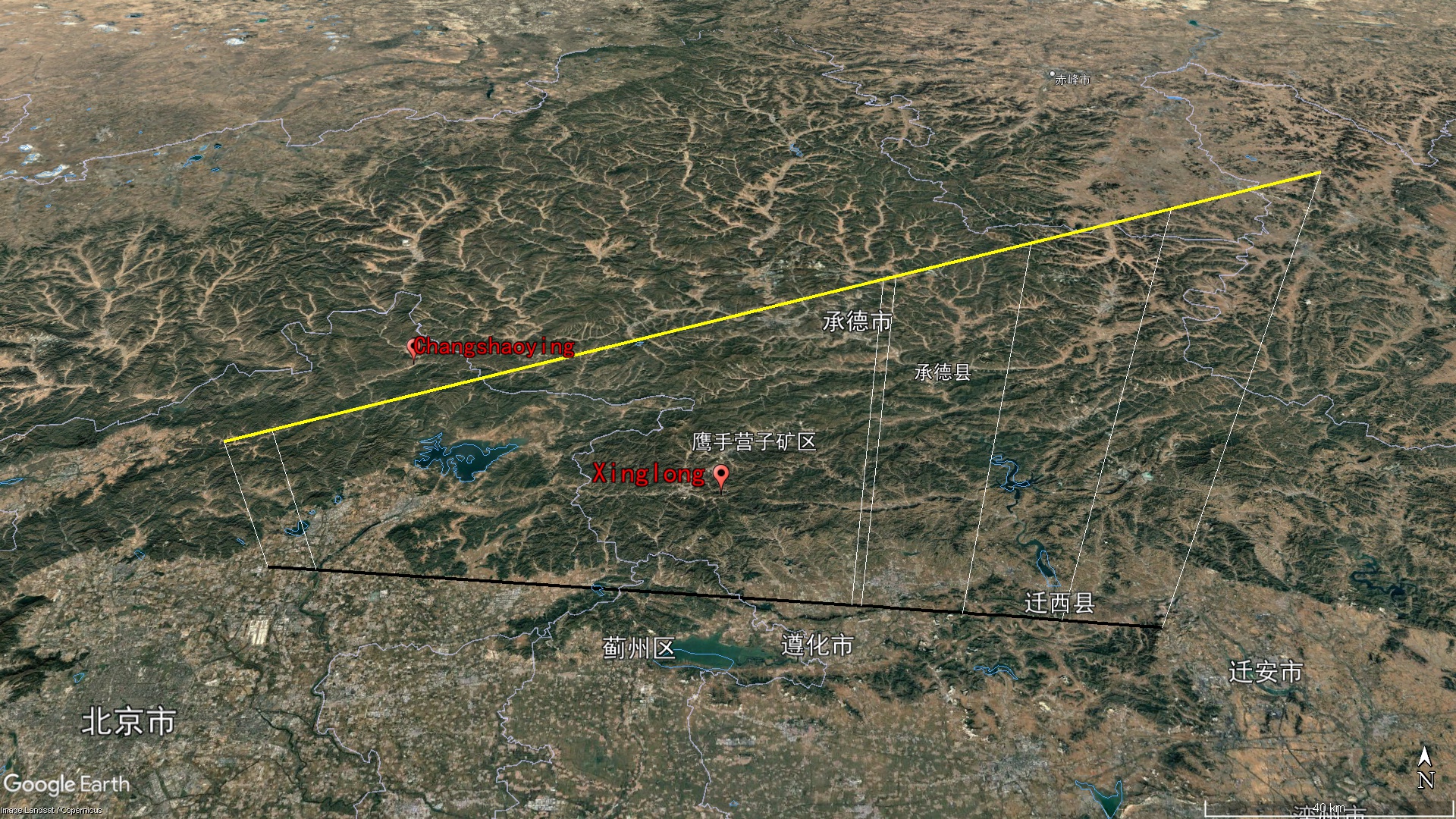}
    \caption{The 3-D trajectory of the meteor occurred on Jan 29th, 2024.
    }
    \label{fig:0129traj}
\end{figure}

\begin{figure}[!htb]
    \centering
    \includegraphics[width=0.8\textwidth]{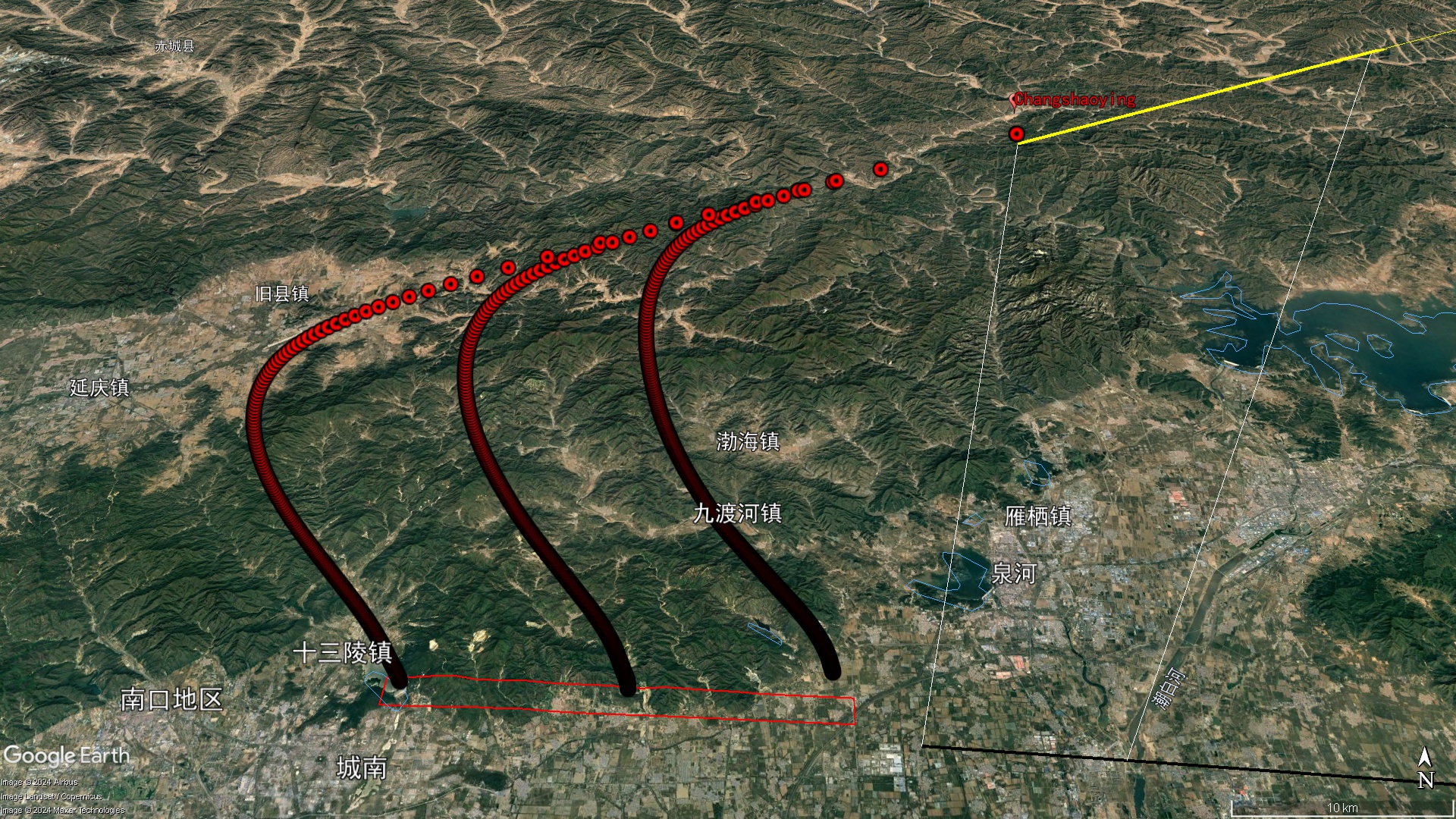}
    \caption{The projected dark flight trajectories of different mass assumptions, which are 1kg, 0.1kg and 0.01kg, from left (west) to right (east), respectively.
    }
    \label{fig:falling}
\end{figure}

The flight path of the meteoroid, as shown in Fig. \ref{fig:0129traj}, is 274 degrees from the east and had a pitching angle of $-$19 degrees. The brightest fragment, as labeled in Fig. \ref{fig:0129image}, disappeared at an altitude of 31.5 km with a speed of 9 km/s. Based on the observational results, a simulation was made to find out the possible landing sites of the meteorites using the software \textit{Meteorite Finder} \citep{CARBOGNANI2024115845}. Three different masses were assumed: 1 kg, 0.1 kg, and 0.01 kg. The results show that the meteorites were likely to land at some of the villages in Changping District, as shown in Fig. \ref{fig:falling}.


These simulations took into account the wind drift using meteorological data. These data were obtained by a weather balloon, which was routinely launched by Beijing Meteorological Observatory at 20:00 Beijing time, 20 minutes after the fireball event. There was the west wind at high altitude and the south wind at low altitude. Thus, the landing predictions drifted to the northeast. Lighter meteorites were more affected. The predicted landing sites are primarily residential and rural areas at the outskirts of Beijing city. Searching for the meteorites was coordinated at these areas, but no meteorites have yet been found.

\section{Summary} \label{sec:summary}

Meteor phenomena are deeply related with the environment of the planetary space. Optical observations for different locations can determine the trajectories of the meteoroids, and subsequently help the study of their origin and evolution. Continuous meteor monitoring is essential for discovering new meteoroid streams and studying of the distribution of planetary dusts. To monitor the meteor activities, we designed and built a meteor monitoring system. The system uses CMOS cameras to cover the full sky. Meanwhile, the accompanied software package can detect the meteors, and measure their positions for determining their orbits. 

In this paper, we present the upgrade and deployment of our meteor monitoring system. The software of the system is better integrated to streamline data processing and transferring. The image data of detected meteors are transferred directly to the measuring pipeline, reducing the total time of getting the celestial coordinates of meteors. Different data products are provided, including FITS files for uncompressed image data, JSON for coordinate data, and PNG and MP4 files for displaying. Other functionalities are provided to expand the capabilities of the software kit.

The software system running at the stations is connected to the data platform via WebSocket and REST requests. The station can upload status report and meteor data, and it can also receive configurations and commands using the connection. This enables the remote and automatic management of many stations.

The data platform is responsible for communication with stations, and it can determine the orbits of the meteors once their coordinate data are received. It uses multiple cloud infrastructures to ensure reliable and scalable services. Minimal changes are required when the amount of meteor increases during major meteor showers and the network expands.

The system has been tested for a year with 4 continuously operating stations. During the operation, 8,683 orbits are obtained, of which about half are from 265 known meteoroid streams. Most of the orbits can be obtained within 1 minute. The detection limit of meteors is $\sim$1 magnitude.

Preliminary analysis was taken with the first year data. The populations of the radiants of the meteoroids are visible and coincide with previous studies. In addition, the bimodal distribution of the sporadic meteors indicates their different origins. Two populations with $T_j$ above or below 2 are also different at their distribution of velocities. The uncertainties of the meteor trajectories are estimated using the Monte Carlo method. The results indicate a significant improvement, which can be attributed to the use of USB cameras that generate raw image data along with precise the GPS timestamps.

In January 2024, a bright fireball was detected by the network in Beijing. The trajectory of the meteoroid was then determined by the observational data of three stations. Further analysis of the falling trajectory including aerodynamics and real-time meteorological data has predicted its possible landing positions. Searching efforts of the possible meteorites have been carried out.

In December 2023, 14 stations are installed, increasing the coverage and detection rate of the network. In the future, the system is planned to be further expanded to cover a large portion of China and will act as an important piece of the global effort of continuously monitoring meteor activities.

\subsection {Acknowledgments}

We are indebted to the referee for thoughtful comments and insightful suggestions that improved this paper greatly.

Hu Zou acknowledges the support from Beijing Municipal Natural Science Foundation (grant No. 1222028), the National Key R\&D Program of China (grant Nos. 2023YFA1607800, 2023YFA1607804, 2022YFA1602902, 2023YFA1608100, 2023YFF0714800, and 2023YFA1608303), the National Natural Science Foundation of China (NSFC; grant Nos. 12120101003, 12373010, 12173051, and 12233008), the science research grants from the China Manned Space Project with Nos. CMS-CSST-2021-A02 and CMS-CSST-2021-A04, and the Strategic Priority Research Program of the Chinese Academy of Sciences with Grant Nos. XDB0550100 and XDB0550000.

Jifeng Liu acknowledges support from the NSFC through grant Nos. of 11988101 and 11933004, and support from the New Cornerstone Science Foundation through the New Cornerstone Investigator Program and the XPLORER PRIZE.

We would like to thank the people who offered great help to the installation and maintenance of the stations in Beijing: Xiaoming Teng and Pengfei Liu at Xinglong station, Haidong Wang, Hong Wang and Yu Mao at Changshaoying station, Xiaolong Liu and Jiawei Zhang at Baihuashan station, Dr. Zhengqun Hu and Xinghua Ma at Wuqing station.

We would also like to thank Prof. Jun Cui and Prof. Xiaoshu Wu of Sun Yat-sen University for their pivotal contributions to the deployment of the stations in Guangdong.

\bibliography{cite}

\end{document}